\documentclass[prd,preprint,tightenlines,floatfix,showpacs,preprintnumbers,nofootinbib,eqsecnum,superscriptaddress]{revtex4-1}

\usepackage[utf8]{inputenc}
\usepackage[T1]{fontenc}
\usepackage{lmodern}
\usepackage[english]{babel}

\usepackage{amsbsy}     
\usepackage{amsfonts}       
\usepackage{amsmath}        
\usepackage{amssymb}        
\usepackage{color}

\ifx\pdfoutput\undefined
\usepackage{graphicx}
\else
\usepackage[pdftex]{graphicx}
\usepackage{epstopdf}
\fi

\definecolor{coldw}{rgb}{0,0.4,0}

\definecolor{colfaded}{rgb}{0.75,0.75,0.75}

\begin{document}


\vspace{1cm}

\title{Diffractive $W^\pm$ production at hadron colliders as a \\ test of colour singlet exchange mechanisms}

\author{Gunnar Ingelman}
\email{Gunnar.Ingelman@physics.uu.se} \affiliation{Department of
Physics and Astronomy, Uppsala University, Box 516, SE-751 20
Uppsala, Sweden}
\author{Roman Pasechnik}
\email{Roman.Pasechnik@thep.lu.se}
\affiliation{Department of Astronomy and Theoretical Physics, Lund
University, S{\"o}lvegatan 14A, SE-223 62 Lund, Sweden}
\author{Johan Rathsman}
\email{Johan.Rathsman@thep.lu.se}
\affiliation{Department of Astronomy and Theoretical Physics, Lund
University, S{\"o}lvegatan 14A, SE-223 62 Lund, Sweden}
\author{Dominik Werder}
\email{Dominik.Werder@physics.uu.se} \affiliation{Department of
Physics and Astronomy, Uppsala University, Box 516, SE-751 20
Uppsala, Sweden}

\begin{abstract}

We revisit diffractive and exclusive $W^\pm X$ production at hadron
colliders in different models for soft colour exchanges. The process
$pp\to p[W^\pm X]p$, and in particular a $W^\pm$ charge asymmetry,
has been suggested as a way to discriminate diffractive processes as
being due to pomeron exchange in Regge phenomenology or QCD-based
colour reconnection models. Our detailed analysis of the latter
models at LHC energies shows, however, that they give similar
results as pomeron models for very leading protons and central
$W^{\pm}X$ production, including a vanishing $W^\pm$ charge
asymmetry. We demonstrate that soft colour exchange models provide a
continuous transition from diffractive to inelastic processes and
thereby include the intrinsic asymmetry of inelastic interactions
while being at the same time sensitive to the underlying
hadronisation models. Such sensitivity also concerns the
differential distributions in proton momentum and $W^\pm$ transverse
momentum which opens possibilities to discriminate between different
colour reconnection models.
\end{abstract}

\maketitle

\section{Introduction}
Diffractive and exclusive processes in Quantum Chromodynamics (QCD)
still remain a theoretically unsolved and intriguing chapter of the
Standard Model of particle physics. Considerable progress has been
made in recent years by focusing on diffractive hard scattering
processes~\cite{Ingelman:1984ns}, where a hard scale defines a
partonic subprocess which can be calculated perturbatively and used
as a well-defined back-bone for the poorly understood soft processes
that give rise to the characteristic features of diffraction in
terms of a leading proton or a large gap in rapidity with no
particle production. In such processes the dominating effect is thus
caused by soft fluctuations of the gluonic field at large distances
making diffractive observables very sensitive to non-perturbative
QCD dynamics and, thereby, providing a tool to explore this unsolved
sector of QCD.

Considering as low scales as $\mu_{\rm soft}\sim
\Lambda_{\rm{QCD}}$, individual gluons are not resolved and one
should rather consider collective gluon fields, such as modeled by
colour string-fields in the Lund hadronisation
model~\cite{Andersson:1983ia}, or even hadron-like objects, such
as modeled through pomeron exchange in the Regge
approach~\cite{Polkinghorne:1980mk,Forshaw:1997dc}. This
has led to different approaches to describe the soft dynamics of
diffractive processes: on the one hand, models based on pomeron
exchange using Regge phenomenology initially developed in the pre-QCD era and,
on the other hand, models based on soft gluon exchange between
hard-scattered partons and beam hadron remnants, which can modify
the colour topology between the emerging partons resulting in a
different final state of hadrons, e.g. with rapidity gaps. The
latter type of dynamics was first introduced in the Soft Colour
Interaction (SCI) model~\cite{Edin:1995gi} and has later been
developed in various ways such as the Generalized Area Law (GAL)
model~\cite{Rathsman:1998tp} making the probability for colour
exchanges dynamical.

Many different diffractive hard scattering processes have been
observed experimentally and studied theoretically
\cite{Ingelman:2005ku}. Much attention has been given to central
exclusive processes~\cite{Albrow:2010yb}, in particular, the
spectacular Higgs boson production process $pp\to pHp$ at LHC, where
the Higgs boson mass might be reconstructed from a measurement of
the leading proton momenta~\cite{KMR_Higgs,Khoze:2000jm}. The
estimated cross-section is, however, small and has a substantial
uncertainty due to its dependence on soft QCD
dynamics~\cite{Dechambre:2011py}.

On the experimental side, both the CDF and D0 collaborations at the
Fermilab Tevatron have reported the measurement of several different
diffractive processes~\cite{Abe:1997jp,Abazov:2003ti,Abazov:2010bk,Aaltonen:2012th}.
Of special interest here is the diffractive
gauge boson production for which the CDF experiment recently
reported results based on the forward spectrometer to detect leading
anti-protons~\cite{Aaltonen:2010qe}. Compared to measurements based
on rapidity gaps, this has the advantage of much smaller dependence
on the gap survival and gap acceptance factors,
resulting in more stringent tests of diffractive models.

On the theoretical side, the diffractive production of gauge bosons
has also received attention~\cite{GolecBiernat:2009pj,GolecBiernat:2011dz}
due to a quite
high sensitivity to the production mechanism and at the same time a
large enough cross section to be experimentally observed and studied
in detail. The intricate mechanism of QCD factorisation breaking in
diffractive Drell-Yan and $W,\,Z$ production~\cite{Pasechnik:2011nw}
enhances the interest for this kind of processes.

In this paper, spurred by these recent developments, we will revisit
the SCI and GAL models for diffractive W production at hadron
colliders. After a short recapitulation of the essence of these
models we will compare with the most recent data on leading
antiprotons from the Tevatron and make predictions for double
leading protons at the LHC. In particular, we will clarify the
recent claim \cite{GolecBiernat:2011dz} on $W$ charge asymmetry in the latter case.

\section{Colour singlet exchange models}

\begin{figure}[t]
\includegraphics[width=0.7\textwidth]{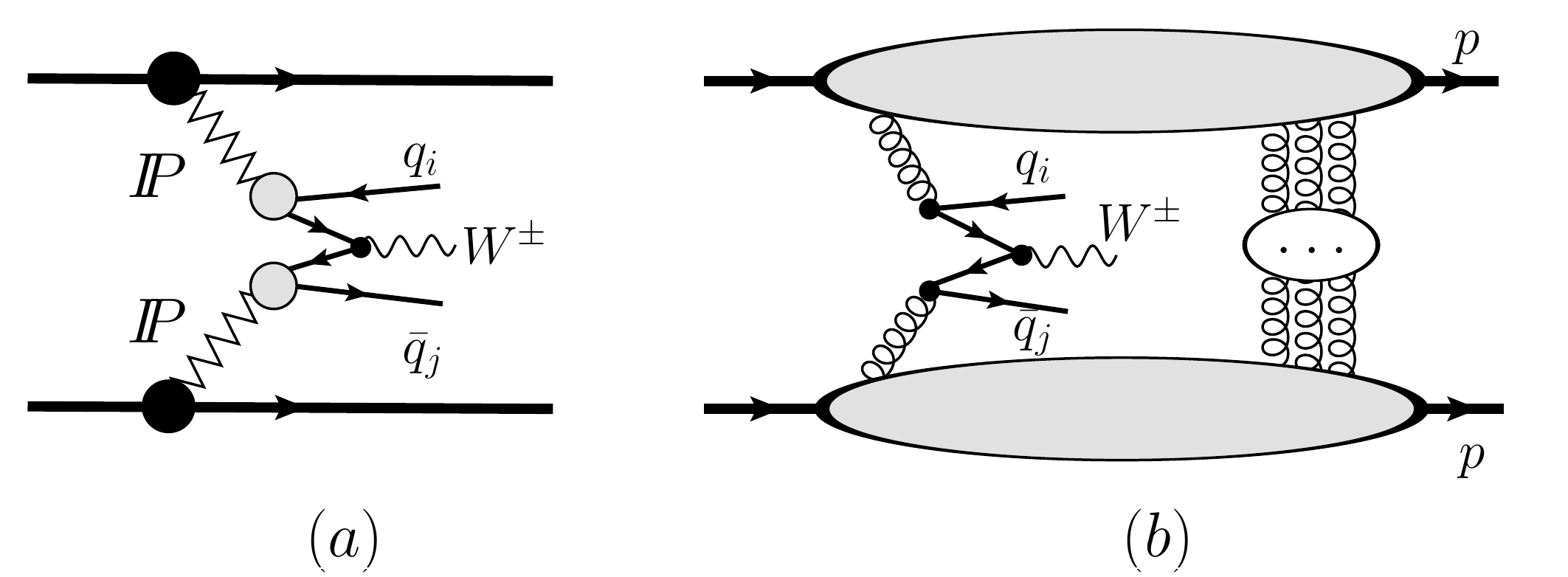}
\caption{The exclusive diffractive process $pp\to p[W^{\pm}X]p$,
with central $W^{\pm} + 2~\rm{jets}$ separated from the final
protons, based on (a) double pomeron exchange in Regge approach and
(b) soft colour exchange in QCD.} \label{fig:Wjj}
\end{figure}

The focus of the paper is on diffractive gauge boson production in
hadron collisions. In particular, we will concentrate on the
exclusive process $pp\to p[W^{\pm}X]p$ at the LHC with $\sqrt{s}=14$
TeV but we will also consider single diffractive W production such
as $\bar{p}p\to \bar{p} [W^{\pm}X]$ at the Tevatron with
$\sqrt{s}=1.96$ TeV. Fig.~\ref{fig:Wjj} illustrates the former for a
typical parton level subprocess where $X$ is a pair of quark jets as
an example. This process will be measured in the near future by the
ATLAS experiment using forward spectrometers~\cite{Royon:2010sm},
and different models of diffraction can then be tested.

On general grounds, the requirement of a leading proton (or
anti-proton) in the final state, which is more or less unscathed,
means that the momentum transfer should be soft,
$\sqrt{|t|}\sim \Lambda_{\rm{QCD}}$, and that larger momentum
transfers are exponentially suppressed.
In addition, only a small fraction of the
proton's longitudinal momentum may be lost, such that $1-z\sim
M_{\rm{WX}}/\sqrt{s}\ll 1$ (for the $W^{\pm}X$ system at central
rapidity $y\sim0$) with $z=|p_z|/p_{\rm beam}$ being the momentum
carried by the leading proton compared to the beam energy.

In the Regge approach, this type of processes are described in terms
of single or double pomeron exchange (DPE), Fig.~\ref{fig:Wjj}a,
using a factorisation into a pomeron flux and parton density
functions (PDF) in the pomeron. Such diffractive PDF's have been
fitted to diffractive deep inelastic scattering   data from the H1
and ZEUS experiments at the $ep$ collider HERA. In this way a
consistent description of diffractive deep inelastic scattering can
be obtained~\cite{Ingelman:2005ku}. The problem is that these
diffractive PDF's are not universally applicable for other
diffractive processes. For example, using them to calculate
diffractive hard scattering processes in $p\bar{p}$ collisions one
obtains cross-sections that are an order of magnitude larger than
observed at the Tevatron~\cite{GS}. Although this problem can be
cured by introducing an overall renormalisation through a soft
rapidity gap survival factor depending on the cms energy, it
represents an incompleteness of the double pomeron exchange model in
general.

As an alternative to the pomeron approach, models have been
developed where soft interactions result in different colour
topologies of the confining string-fields, giving different
hadronic final states after hadronisation. In particular, a rapidity
range without a string-field results in an event with a
corresponding rapidity gap.

The Soft Colour Interaction (SCI) model~\cite{Edin:1995gi} is based
on the exchange of soft gluons, below the conventional cut-off
$Q_0\sim 1$ GeV for perturbative QCD. The momentum exchange does
then not significantly change the momenta of emerging partons, but
the exchange of colour does change the colour structure of the
emerging parton system, resulting in a modified string-field
topology and thereby affecting the resulting distribution of final
state hadrons. In effect, the SCI model introduces a probability,
given by a parameter $P_{\mathrm{SCI}}$, for the exchange of a
colour octet between any pair of partons (including beam and target
spectators) emerging from the perturbative QCD treatment of the
event in the Monte Carlo event generators {\sc
Lepto}~\cite{Ingelman:1996mq} for deep inelastic lepton-nucleon
scattering or {\sc Pythia} \cite{Sjostrand:2006za} for hadron-hadron
scattering. As a result, a modified string topology is  obtained
before the conventional Lund hadronisation
model~\cite{Andersson:1983ia} is applied. In spite of its simplicity
and with a single value
 of the only new parameter
$P_{\mathrm{SCI}}$, this provides a phenomenologically successful
model that can account for a large variety of diffractive data,
including the diffractive structure function at HERA
\cite{Edin:1995gi}, diffractive jets and quarkonia production at the
Tevatron~\cite{Edin:1997zb,Enberg:2001vq}. The model has also been
applied for predicting diffractive Higgs production at the LHC
\cite{Enberg:2002id}.
In the following we will be using the canonical value $P_{\mathrm{SCI}}=0.5$.

In the same spirit as the original SCI model, but with a different
mechanism for non-perturbative colour rearrangements, the
Generalized Area Law (GAL) model has been developed
in~\cite{Rathsman:1998tp}. The GAL model was a first attempt to
make the colour reconnection probability dynamical instead of static
as in the SCI model. In short it employs the difference in
generalized string area for two different string configurations to
weight the reconnection probability, $ P_{\rm GAL} = P_0
\left[1-\exp(-b \Delta A) \right]$, where $P_0\sim 0.1$ is the
maximal reconnection probability of order $1/N_C^2$, $b$ is the
string parameter (PARJ(42) in {\sc Pythia}), and  the area
difference is defined as $\Delta A = A^{\rm old}- A^{\rm new}$ with
the area for a string piece between partons $i$ and $j$ being
$A(p_i,p_j)=2(p_i\cdot p_j - m_i\cdot m_j)$.
We will use the standard value $P_{0}=0.1$.
The model has been
shown to give a good description of the diffractive structure
function at HERA~\cite{Rathsman:1998tp} as well as other
characteristics of both the diffractive and inclusive final
state~\cite{Edin:1999jq}. Both the SCI and GAL models have recently
been adapted to {\sc Pythia} 6.4~\cite{SCIGALPythia}.

Although formulated in terms of interactions or rearrangements of
strings, the GAL model describes the transition from a parton state
with a given colour configuration at the scale $Q_0$ to a set of
strings at the soft scale $\mu_{\rm soft}\sim \Lambda_{\rm{QCD}}$.
The SCI model, on the other hand, is formulated in terms of exchange
of gluons, although softer than the factorized dominating hard
partonic interactions, they may have scales anywhere in the range
from such a factorisation scale down to the hadronisation scale,
$\mu_{\rm soft}\sim \Lambda_{\rm{QCD}}$. Even if considering a
factorisation scale as low as the perturbative QCD cut-off $Q_0\sim
1$ GeV, this range is not small in the logarithmic measure
applicable in QCD. Therefore, significant soft colour exchanges are
to be expected --- the problem is how to properly describe them. A
theoretical QCD-basis for SCI-like models has been proposed
in~\cite{Brodsky:2004hi} and developed into a dynamical colour
exchange model later in~\cite{Pasechnik:2010zs}.

The common feature of the various colour reconnection models is that
the hard production process of the $ [W^{\pm}X]$ system is described
using standard collinear factorisation. Given the requirement of
leading protons, the momentum fractions of the initiating partons
will be $x_1\sim x_2 \sim M_{\rm{WX}}/\sqrt{s}\ll 1$ for the $W^{\pm}X$
system at central rapidity $y\sim0$. Such small-$x$, processes are
expected to be dominated by gluons, e.g.\ $gg\to Wq{\bar q}$, due to
the large gluon density. The quark (sea or valence) content of the
proton could become noticeable at larger $x_1$ or $x_2$, i.e. at
larger rapidities or larger $M_{\rm{WX}}$. In this case, one or even
both protons will predominantly be destroyed by the interaction,
giving a reduced contribution to the process of interest.

As illustrated in Fig.~\ref{fig:Wjj}b, the colour octet charge of
the gluon initiating the hard subprocess can be compensated by
additional gluon exchanges resulting in an overall colour singlet
exchange. Provided that all these gluons have small transverse
momenta, the proton can remain in a coherent state with only some
loss of longitudinal momentum. In accordance with the uncertainty
relation, the time and space ordering of these exchanges cannot be
specified better than the inverse of the momentum transfer scale.
When this is soft, colour exchange is possible before the coherence
of the proton is destroyed. When it stays intact
one has the leading proton characteristic for diffractive
scattering. When it does not stay intact, it may emerge as an
excited small mass state. In case the gluon exchange does not
constitute a singlet exchange, a colour charged proton remnant will
emerge and hadronise, which may still produce a leading proton, but
then at a lower fraction of the initial beam momentum.

This scenario is compatible with the exchanged gluons as parts of
the bound state proton and given by standard parameterisations of
the gluon density $g(x,Q_0^2)$, giving a large probability for
gluons with small longitudinal momentum fraction $x$. Moreover, with
the conventional gluon $p_{\perp}$-distribution at $Q_0$ given by a
Gaussian distribution of width $\sim \Lambda_{\rm{QCD}}$ (often
referred to as intrinsic $p_{\perp}$), one naturally obtains the
experimentally observed distribution $e^{-bt}$, with $b\sim
1/2\Lambda_{\rm{QCD}}^2$, of the momentum transfer squared $t\sim
-p_\perp^2/z$ to the final proton.

To summarize, the common key feature of diffractive-like events
generated by colour reconnection models, is the dominance of a
gluon-initiated hard parton process augmented by additional softer
colour octet exchanges, resulting in a $t$-channel exchange which is
colour singlet and electrically neutral. Thus, the overall
expectation is that no charge asymmetry between diffractive $W^+ X$
and $W^- X$ production should appear, which is contrary to the claim
in~\cite{GolecBiernat:2011dz}.

For diffraction and leading protons more generally, the most
essential issue is how the proton remnant is treated. Conventional
Monte Carlo event generators employ hadronisation models based on
colour triplet string fields, most notably the Lund model
\cite{Andersson:1983ia}. Gluons are here represented by
energy-momentum carrying kinks on a string, but quarks, antiquarks
and diquarks are triplet charges at the end of strings. Therefore, a
colour octet $uud$ remnant is conventionally ``split'' into a quark
and a diquark with triplet and anti-triplet colour charges and
separate four-vectors, which is called a ``cluster'' in the
following. This split occurs even if the above soft colour exchanges
restore the remnant to a colour singlet. In this case the valence
quark and diquark have to be recombined during hadronisation to form
a hadron or a small-mass system which further decays, and the
details of these splitting and recombination procedures affect the
spectrum of diffractive-like leading protons. A proper treatment of
diffraction may motivate a changed Monte Carlo procedure where the
proton's $uud$ remnant is kept as a single object during the
coherence time and only split if it emerges, after colour exchanges,
in a colour octet state.

When the proton remnant is not in a colour singlet state after
colour reconnections, hadronisation can still produce a spectrum of
leading protons,
extending to large fractions $z$ of the beam momentum.

\section{Results}

The following results are obtained by simulations of $pp\to
W^{\pm}X$ events at the LHC energy $\sqrt{s}=14$ TeV as well as
$\bar{p}p\to W^{\pm}X$ events at the Tevatron energy $\sqrt{s}=1.96$
TeV using {\sc Pythia} \cite{Sjostrand:2006za}, with basic hard
subprocess $q\bar{q} \to W$. An implementation of the SCI and GAL
model \cite{SCIGALPythia} for the colour exchanges before
hadronisation with the standard Lund model \cite{Andersson:1983ia}
is used for generating the diffractive events. However, details in
the Monte-Carlo modeling such as the multi-parton interactions and
the treatment of the proton remnants are also crucial for the
resulting leading proton spectrum, as we will demonstrate by
comparing different versions and tunes of {\sc Pythia}. As baseline
we use {\sc Pythia} version 6.425 with the Perugia 0
tune~\cite{Skands:2010ak}, which mainly has been adjusted to data
from the Tevatron. In the following we will start by exploring the
single leading proton spectra at LHC energies. We will then turn to
the rapidity distributions of the $W$'s both at the Tevatron and the
LHC. Finally, we will discuss the question of the $W$ charge
asymmetry.

\subsection{Single leading protons}

\begin{figure}[!t]
\includegraphics[width=0.7\columnwidth]{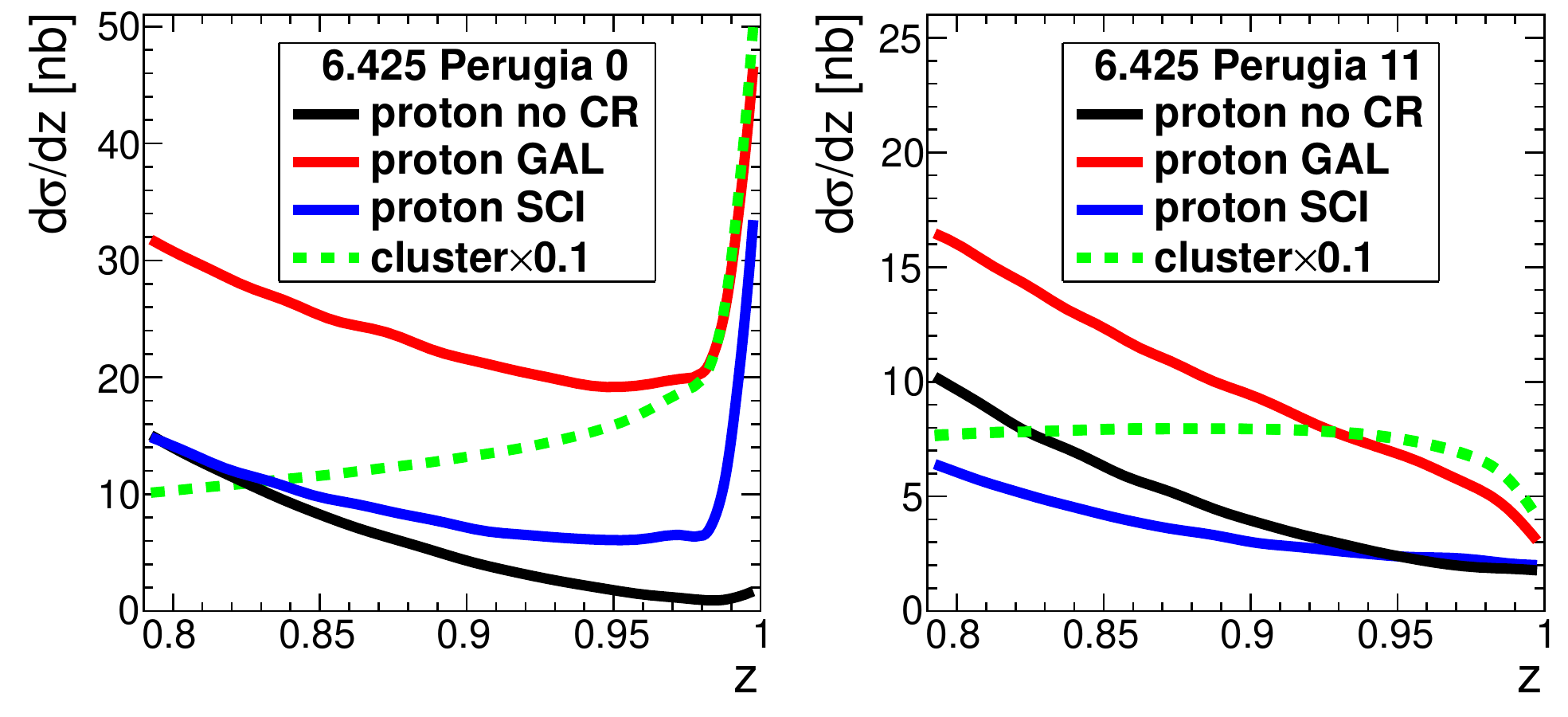}
\includegraphics[width=0.7\columnwidth]{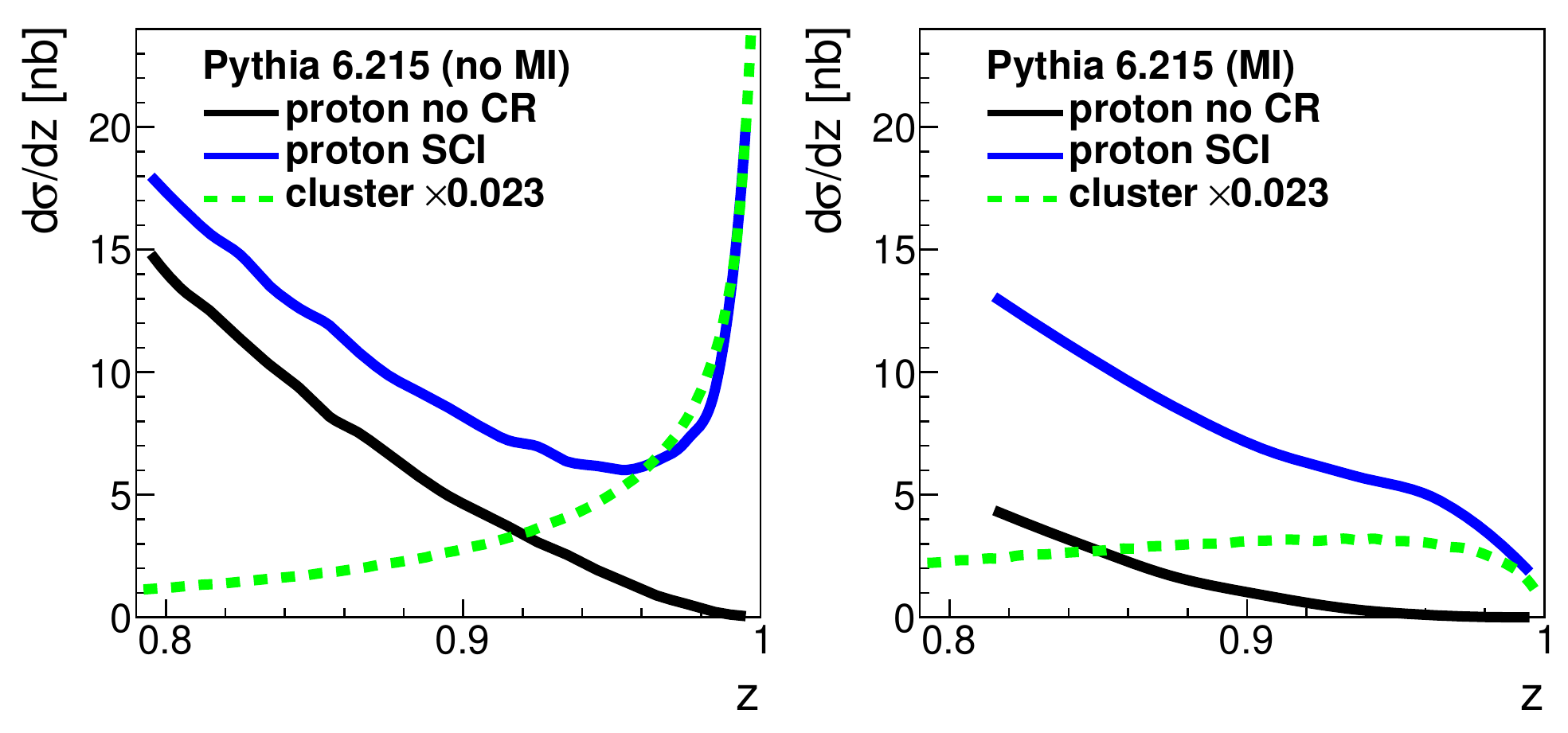}
\caption{ Distribution in momentum fraction $z=|p_z|/p_{\rm beam}$
of the single leading proton in $pp\to p[W^{\pm}X]$ events at
$\sqrt{s}=14$ TeV  obtained from different versions and tunes of
{\sc Pythia} without colour reconnections and with GAL and SCI
models. Leading clusters with $m_{cl}<1.5$~GeV and proton flavor
quantum numbers, but not necessarily colour singlets, are scaled down
to overlap with the diffractive proton peak at $z\to 1$. }
\label{fig:zcomp}
\end{figure}

The basic features of the single leading proton spectrum in
diffractive $W^\pm X$ production at 14 TeV are demonstrated in
Fig.~\ref{fig:zcomp}, showing the momentum distributions of protons
and small mass clusters. The latter are required to have the same
quark content as a proton and invariant masses $m_{cl} \le 1.5$ GeV,
but are not required to be in a colour singlet state. These cluster
spectra have been scaled with a numerical factor such that they
agree with the leading proton spectra for large $z$. The colour
exchange mechanism (SCI or GAL) can turn these clusters into colour
singlet states, giving rise to leading protons after hadronisation.
At the same time the actual amount of leading protons will depend on
the hadronisation mechanism used in the Monte Carlo.
If the cluster mass is above the
threshold for two-particle production $m_{cl} \gtrsim m_p + m_\pi$
it will likely give two particles that share the cluster momentum.
This
also means that the resulting leading proton spectrum will be
sensitive to the masses assigned to the quarks and diquarks in the
proton remnant, as will be made more clear below.
Fig.~\ref{fig:zcomp} top-left clearly shows the two contributions to
the proton spectrum which are the diffractive-like peak from
beam protons staying intact after an overall colour singlet exchange,
and the tail of the hadronisation spectrum. We note that the shape
of the cluster spectrum resembles the proton spectrum in the peak
region.
We
also note that although the normalisation is somewhat different, the
shapes of the leading proton spectra obtained with the SCI and GAL
models are very similar.

However, the forward peak may be lost due to details in the Monte
Carlo models. As an example, in the Perugia 11 tune shown in the
top-right corner of Fig.~\ref{fig:zcomp}
there is no ``diffractive peak'' even at parton
level and, hence, also not at hadron level. The reason for this is
that in the Perugia 11 tune dipoles stretched between perturbative
partons and the beam remnant are allowed to radiate in the forward
direction. Not only is it doubtful to what extent one can properly
define dipole radiation from such a system but, in addition, this
effectively means that the non-perturbative remnant is radiating
perturbative partons in contradiction with the leading proton
coherence condition.

For comparison, we add the results for the same observable from the
older {\sc Pythia 6.215} using the old virtuality-ordered parton
shower and underlying event model based on multiple interactions
treated separately from the parton shower at difference to the new
{\sc Pythia} version
where they are intertwined.
More specifically the latter means that there is a common Sudakov form factor
for both intitial and final parton showering as well as the multiple interactions,
instead of one for each.
As a consequence, the exponentially suppressed tail of the
distribution giving events with very low extra activity is different in the two versions
but precisely these events contribute to the diffractive sample.

The bottom row of
Fig.~\ref{fig:zcomp} shows the result of {\sc Pythia} 6.215 with
and without multiple interactions.
Removing additional partons from the proton as is
done by multiple interactions certainly reduces the momentum fraction left
for the
remnant, which may result in smearing out the ``diffractive peak''
and shifting it down to smaller momentum fractions. The resulting
protons are now mixed with the contribution of protons coming from
string hadronisation, which makes it impossible to single out the
``diffractive'' component.

This implementation of multiple interactions was also used
in~\cite{GolecBiernat:2011dz} in the study of the $W$ charge
asymmetry in the SCI model together with a lower cut on the forward
protons of $z^{cut}=0.85$. Based on Fig.~\ref{fig:zcomp}
bottom-right we note that the resulting event sample contains large
contributions from the quark-induced subprocesses, instead of only
charge-symmetric gluon-induced ones, and thereby a non-negligible
source for the $W$ asymmetry from such a non-diffractive sample has
emerged as will be made more clear below. In the
Regge approach, this corresponds to contributions induced not only by
pomeron exchange, but also by Reggeon exchanges in terms of meson
trajectories. These are expected to introduce a charge asymmetry,
since e.g. ``meson''
exchange of $\pi^+$ quantum numbers leaving a
leading neutron is less suppressed than a $\pi^-$ exchange leaving a
more massive forward $\Delta^{++}$. Thus, comparing pomeron exchange
alone with soft colour exchange models can only be done in the peak region
of $z\to 1$ up to hadronisation
corrections discussed above.

It should be noted that the diquark fragmentation tail clearly seen
in Fig.~\ref{fig:zcomp} is inherent to all hadronisation models and
is always there irrespectively of whether one employs a colour
reconnection model or not. It is also clear from the figure that for
large $z\to1$, the leading proton spectra obtained with colour
reconnection models follow the one from the leading clusters. It
is thus natural to use the difference between the leading proton
spectrum with reconnections and the one without them as the genuine
diffractive contribution. At the same time, this simple picture is
complicated by the fact that such leading protons can also arise in
the Monte Carlo from the combination of the valence diquark and a
sea quark with the right quantum numbers. In this case, the
coherence of the proton can clearly not be retained and, therefore,
this should not be considered as part of the diffractive sample. It
is, therefore, not completely clear where one should draw the line
between diffractive and non-diffractive contributions. This is a
natural consequence of the colour reconnection models having no
sharp distinction between these two types of events but instead providing a
smooth transition between diffractive and inclusive processes
\cite{Edin:1995gi}.

\subsection{$W$ charge asymmetries}

Having established these properties of the single leading proton
spectrum in the Monte Carlo model, which are of fundamental importance for any
discussion of diffractive-like phenomena, we now turn to the $W$
charge asymmetries in the case of single and double leading protons.
\begin{figure}[!t]
\includegraphics[width=0.45\columnwidth]{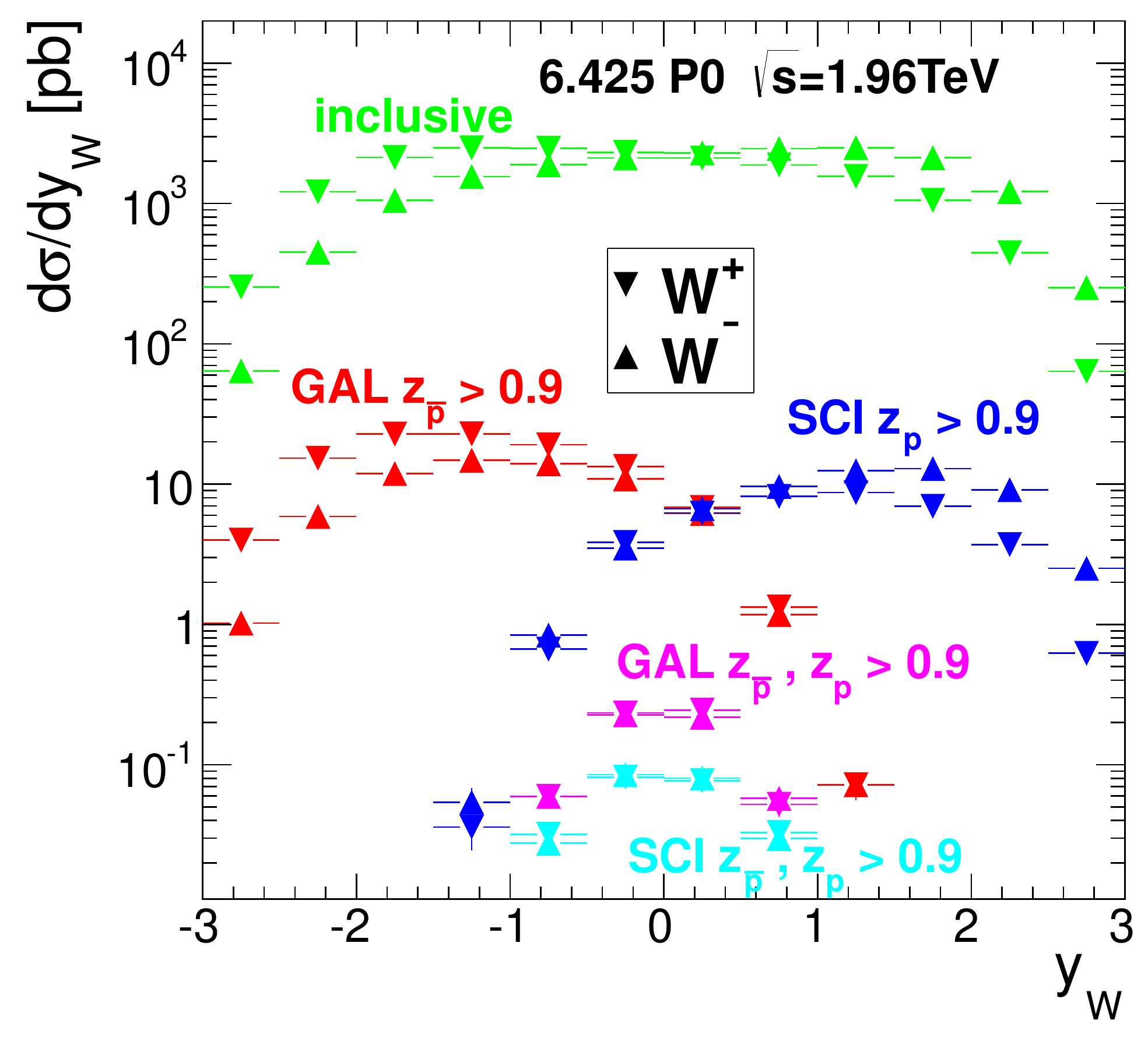}
\includegraphics[width=0.45\columnwidth]{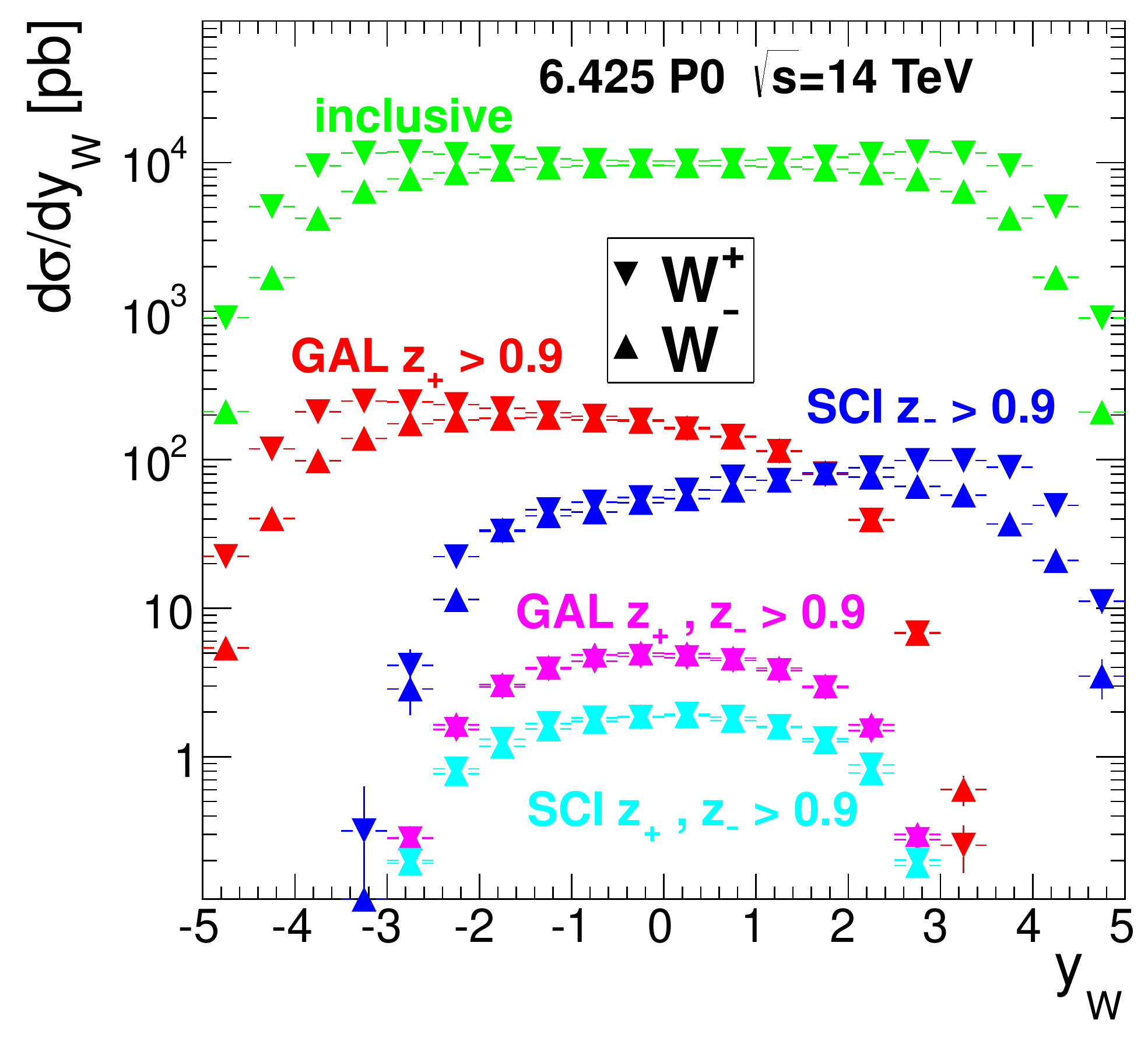}
\caption{ The distribution in rapidity of inclusive $W^\pm$ production
compared to the results when requiring single or double leading protons
in the GAL and SCI models for
the Tevatron (left) and LHC (right) energies, respectively. In the
Tevatron case $z_{\bar{p}}$ ($z_{p}$) denotes the fractional momentum of the
leading antiproton (proton) compared to the beam energy, whereas for LHC
$z_{+}$ ($z_{-})$ is the fractional momentum of the leading proton in
the positive (negative) direction. The results have been obtained with
{\sc Pythia} 6.425 using the Perugia 0 (P0) tune.
}
\label{fig:rapcomp}
\end{figure}

We start by showing the rapidity distributions of the produced $W$'s
when requiring single or double leading protons (or antiproton for
the Tevatron), where a leading proton is operationally defined as
having $z>0.9$. Fig.~\ref{fig:rapcomp} shows the resulting
distributions obtained by using GAL and SCI models comparing also to
the inclusive rate. The left plot shows results for the Tevatron with
the antiproton beam assumed to be along the positive z-axis. As is
clear from the figure, the ratio of the cross-section for having a
single leading antiproton (illustrated for the GAL model) as well as
a single leading proton (shown for the SCI model) to the inclusive
one is close to 1 \% (taking into account a factor two for the
leading protons the ratio is 1.0 \% for GAL and  0.5 \% for SCI)
whereas the ratio of double leading to single leading rates is
smaller and amounts to 0.3 \% for GAL and 0.2 \% for SCI. This can
be compared with the recent results from the CDF experiment at the
Tevatron~\cite{Aaltonen:2010qe}. They find that $(1.00\pm0.11)$\% of
the $W$'s are produced with a single leading proton or antiproton
with $0.90<z<0.97$ and $-1<t<0$ GeV$^2$ and that the fraction of
double leading to single leading protons is less than 1.5\%.
Although the experimental measurements done at the Tevatron are not
precisely for the same conditions, the results are very encouraging,
and the overall agreement is as good as can be expected without
having resorted to retuning of the Monte Carlo model.

Going to LHC energies, as depicted in the right panel of
Fig.~\ref{fig:rapcomp}, the ratio of single leading protons to
inclusive is about 3 \% with the GAL model (again including a factor
two to take into account both sides) to be compared with 1 \% with
the SCI model and the ratio of double leading to single leading is
0.8 \% (0.9 \%  for SCI). From the figure it is also clear that the
higher energy at the LHC opens up a much larger $W$ rapidity region
both when requiring single and double leading protons. In addition,
whereas for the single leading (anti)protons there is an asymmetry
between $W^+$ and $W^-$ very similar to the inclusive one, in the
case of double leading protons any charge asymmetry is much smaller
than the inclusive one. It should be clear that there is an
additional uncertainty in these results due to the extrapolation of
both the colour reconnection and hadronisation models
to LHC energies. However, a detailed analysis of this uncertainty
goes beyond the scope of the present paper.

\begin{figure*}[!t]
\includegraphics[width=0.45\textwidth]{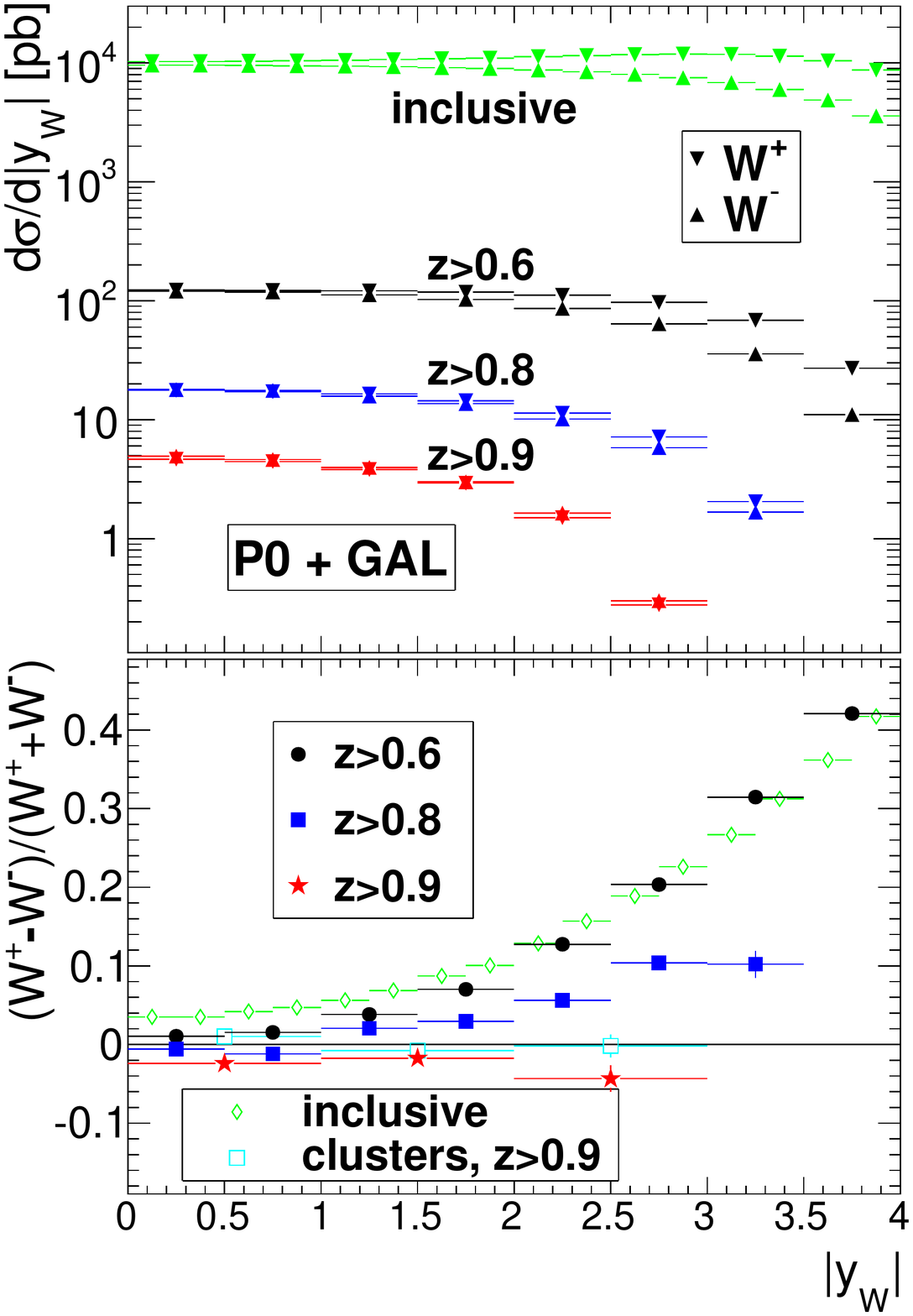}
\includegraphics[width=0.45\textwidth]{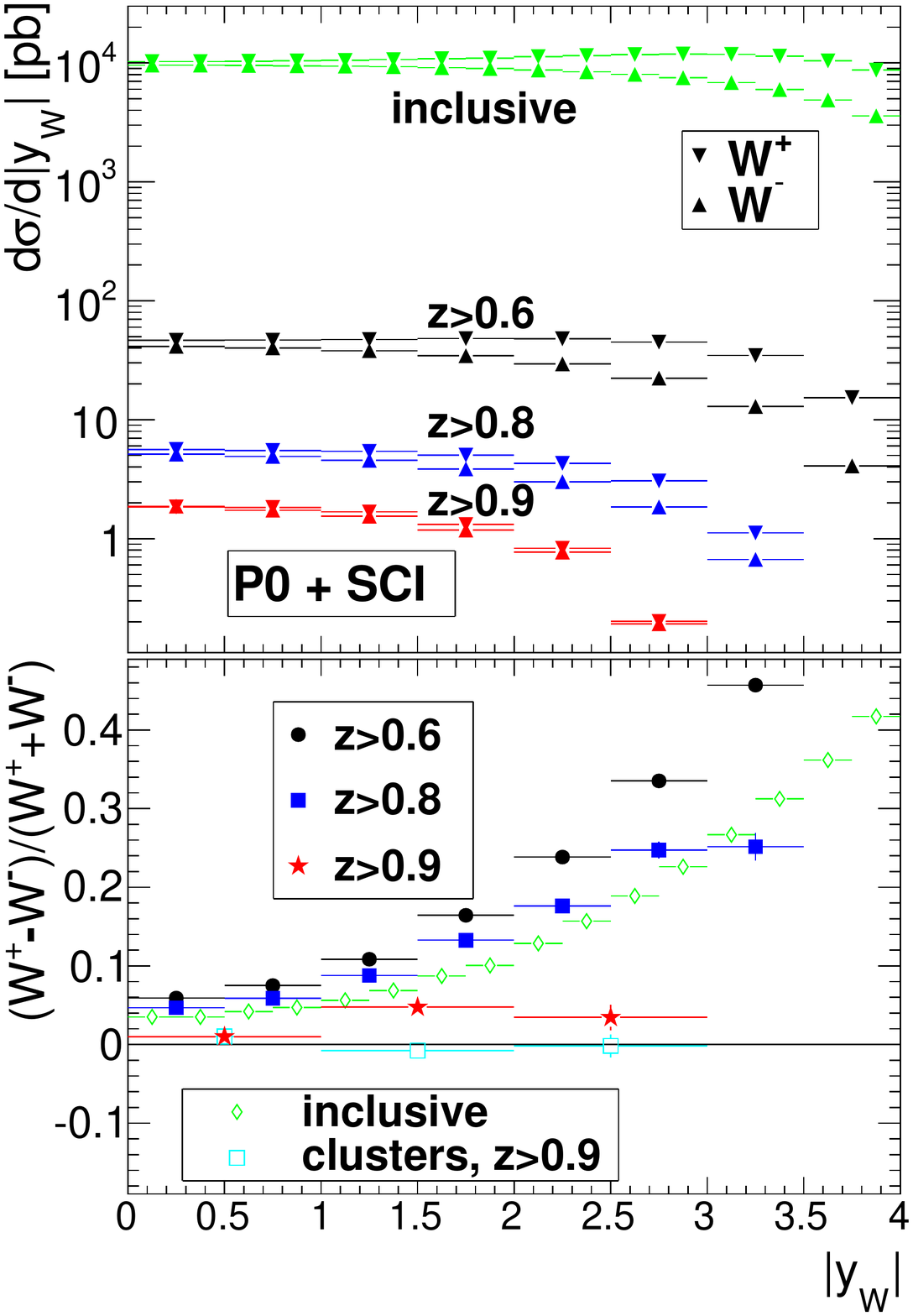}
\caption{ The differential cross sections in rapidity $y_W$ (top)
and the corresponding charge asymmetries (bottom) for the GAL (left)
and SCI (right) models. The curves correspond to the double leading
protons, unless stated otherwise, obtained with {\sc Pythia} 6.425
using the Perugia 0 (P0) tune model.} \label{fig:y-asym}
\end{figure*}

In order to investigate the asymmetries in more detail, we start by
considering the rapidity distributions of $W^\pm$'s and the
corresponding asymmetries at LHC energies in Fig.~\ref{fig:y-asym} for different
cuts on $z$ of the leading protons on both sides and for comparison
the inclusive distributions without any $z$-cut. As can be seen
clearly from the figure, for both the GAL and SCI models
the rates as well as the asymmetries are
strongly dependent on the $z$ cut. For not so strong cuts on $z$ the
asymmetry is close to the inclusive one
whereas for stronger cuts $z\gtrsim0.9$ the asymmetry goes away at the
percentage level. For the GAL model it even becomes slightly negative,
although this may depend on tunable parameters.
To show this we also include a curve with the asymmetries for double
leading clusters with $z>0.9$.

From the figure it is also clear that
for the SCI model the asymmetries are generally larger than for the
GAL model except for $z \to 1$. The reason is that in the SCI model
the leading protons with  $z\lesssim0.9$ are mainly produced from
diquark fragmentation as will become more clear below.
Finally we also see that harder cuts on $z$ correspond
to more central production of the $W^{\pm}$, which is a simple
kinematical consequence of requiring leading protons. For example,
$z>0.9$ means that the cms energy of the $W^{\pm}X$ system is less
than $\sqrt{\hat{s}_{\max}}=1.4$ TeV and thus the rapidity of the
$W^{\pm}$ is limited to $|y_W|<\log \sqrt{\hat{s}_{\max}}/m_W=2.86$.

\begin{figure*}[!t]
\includegraphics[width=0.45\textwidth]{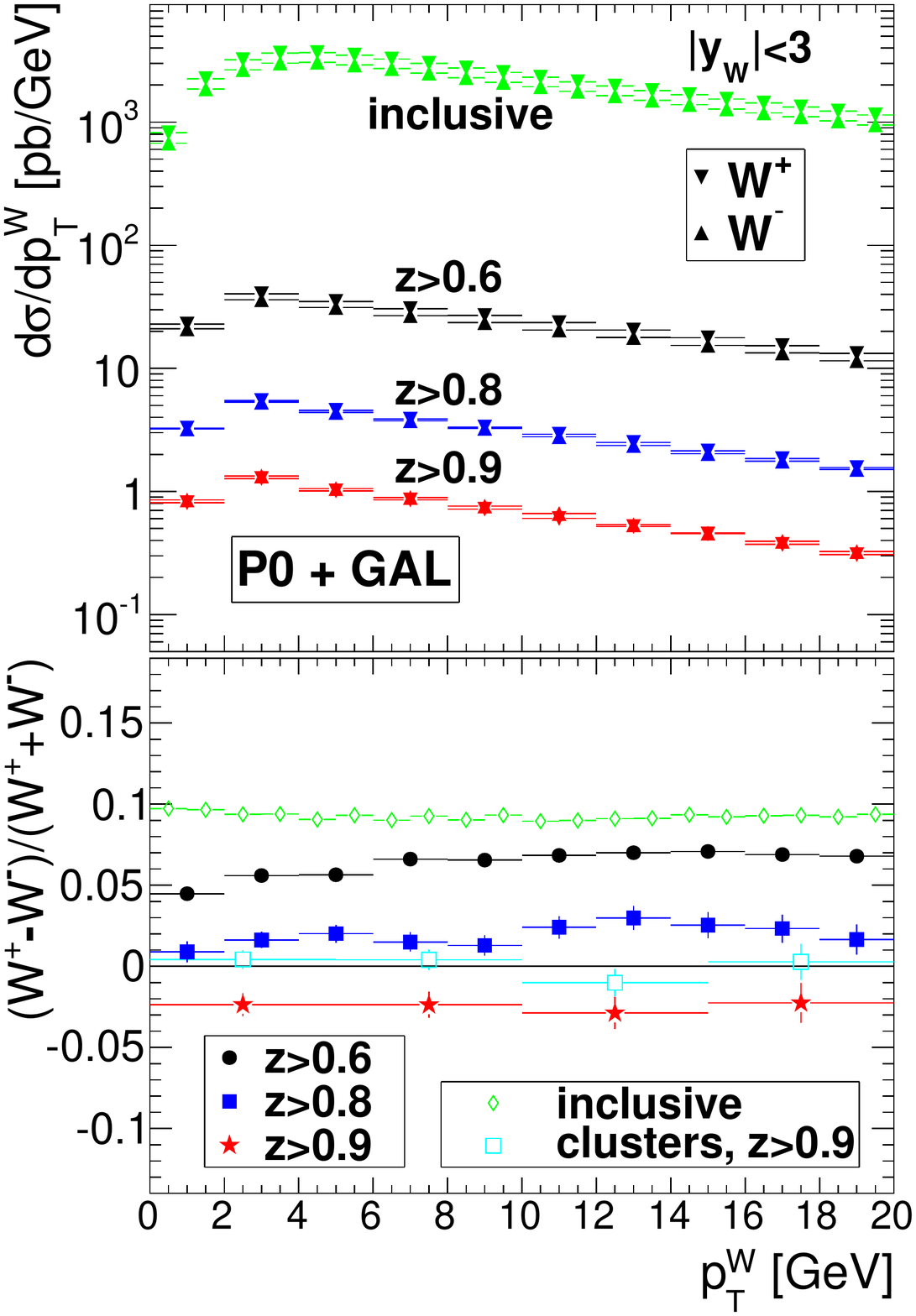}
\includegraphics[width=0.45\textwidth]{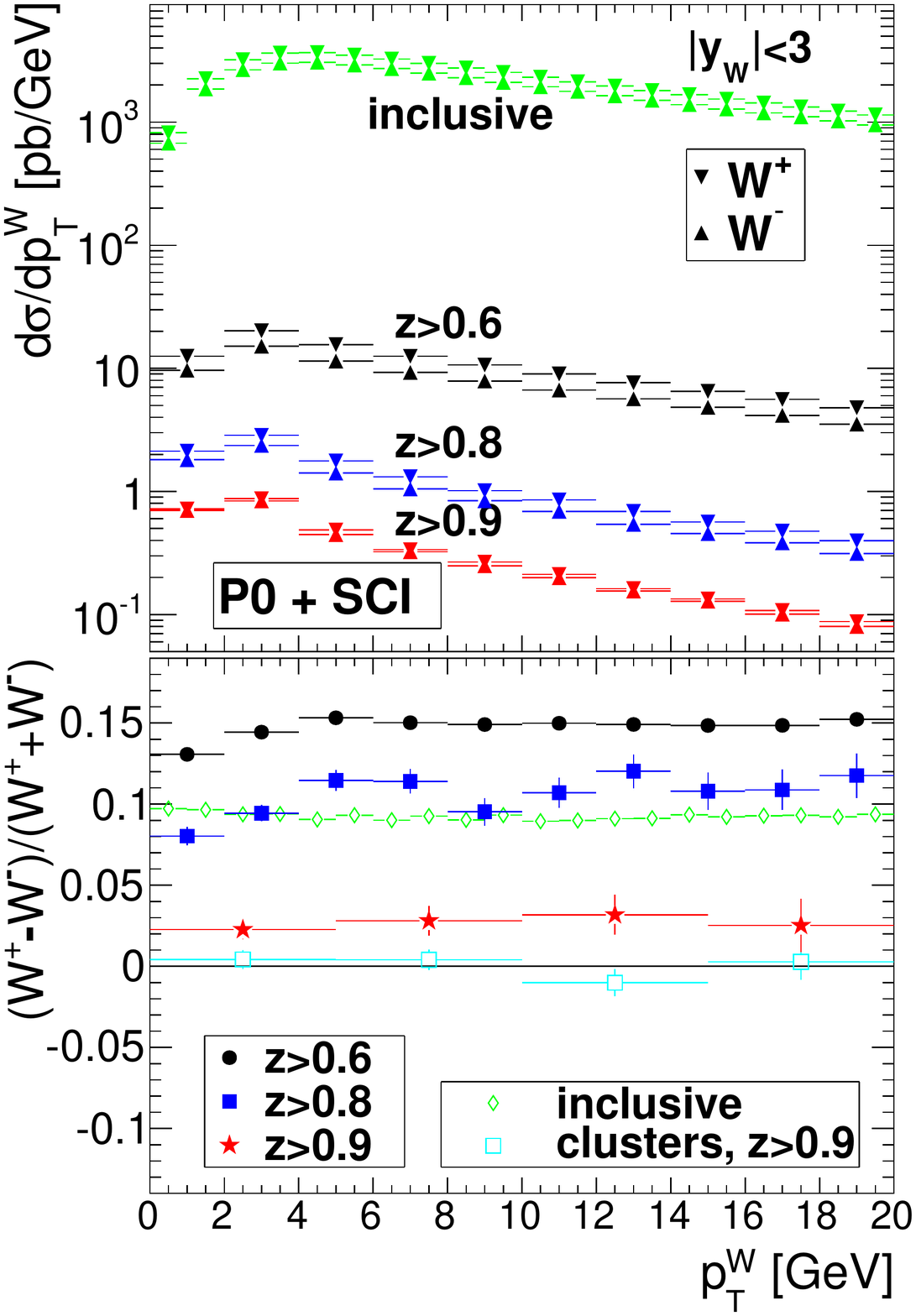}
\caption{ The differential cross sections in transverse momentum  $p_T^W$ (top) and the
corresponding charge asymmetries (bottom) for the GAL (left) and SCI (right) models. The curves correspond to
the double leading protons, unless stated otherwise, obtained with
{\sc Pythia} 6.425 using the Perugia 0  (P0) tune.}
\label{fig:pT-asym}
\end{figure*}

Fig.~\ref{fig:pT-asym}  shows the transverse momentum $p_T$
distribution for the $W^{\pm}$. We first note that for both
the GAL and SCI models the requirement
of double leading protons enhances the cross-section for small $p_T$
compared to the inclusive one, which is natural given the way the
reconnection models are constructed.
We also see that the $W$ $p_T$
spectrum becomes slightly steeper at large $p_T$ when requiring high-$z$
protons from the SCI model, since the increased parton multiplicity in
high-$p_T$ events imply increased combinatorics for soft colour exchanges
that in turn reduces the probability for the proton remnant to emerge
as a colour singlet.
Turning to the charge
asymmetry, it is again clearly visible for the inclusive production,
although mostly as an overall difference in the normalization for
$W^{+}$ and $W^{-}$ respectively. The effects of requiring more and
more leading protons can also be clearly seen giving essentially no
or little asymmetry for $z>0.9$ in both models. The remaining asymmetry is of the
order a few percent and is the result of
hadronisation effects, which
again can be seen comparing to the asymmetry for clusters and is thus
well within an overall uncertainty of the diffractive Monte Carlo modeling.

\begin{figure*}[!t]
\includegraphics[width=0.353\textwidth]{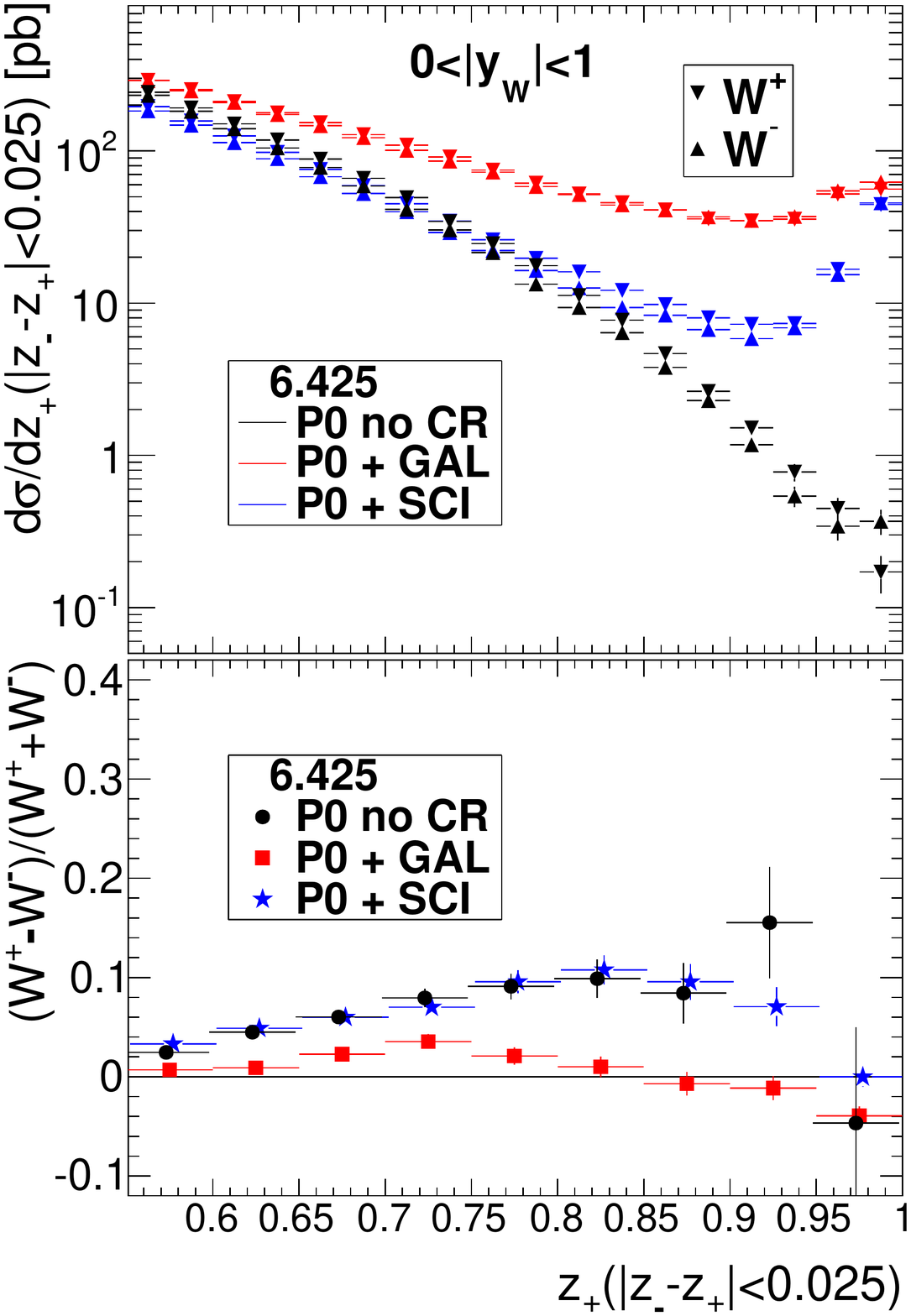}
\hspace*{-0.1cm}
\includegraphics[width=0.30\textwidth]{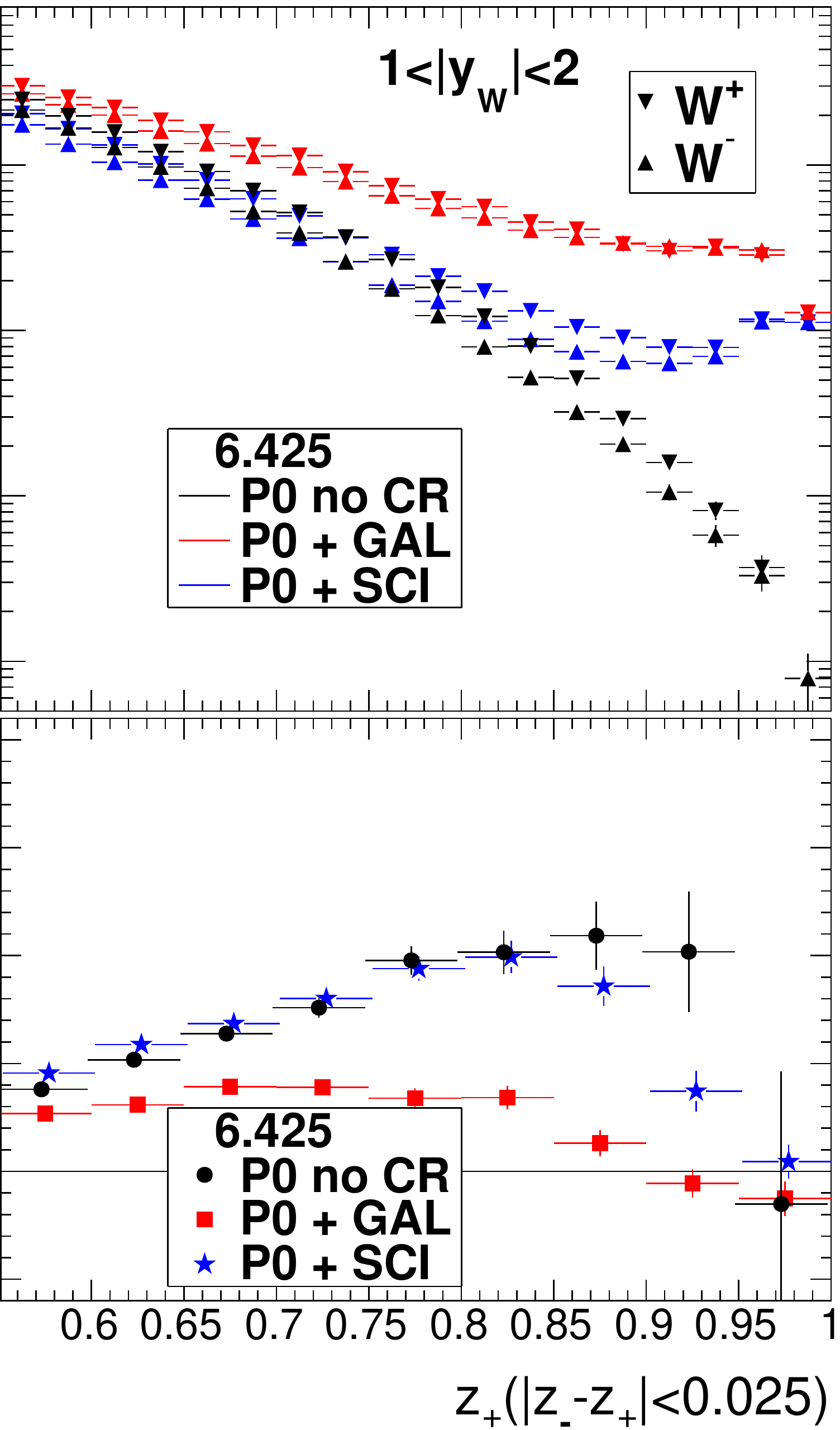}
\includegraphics[width=0.30\textwidth]{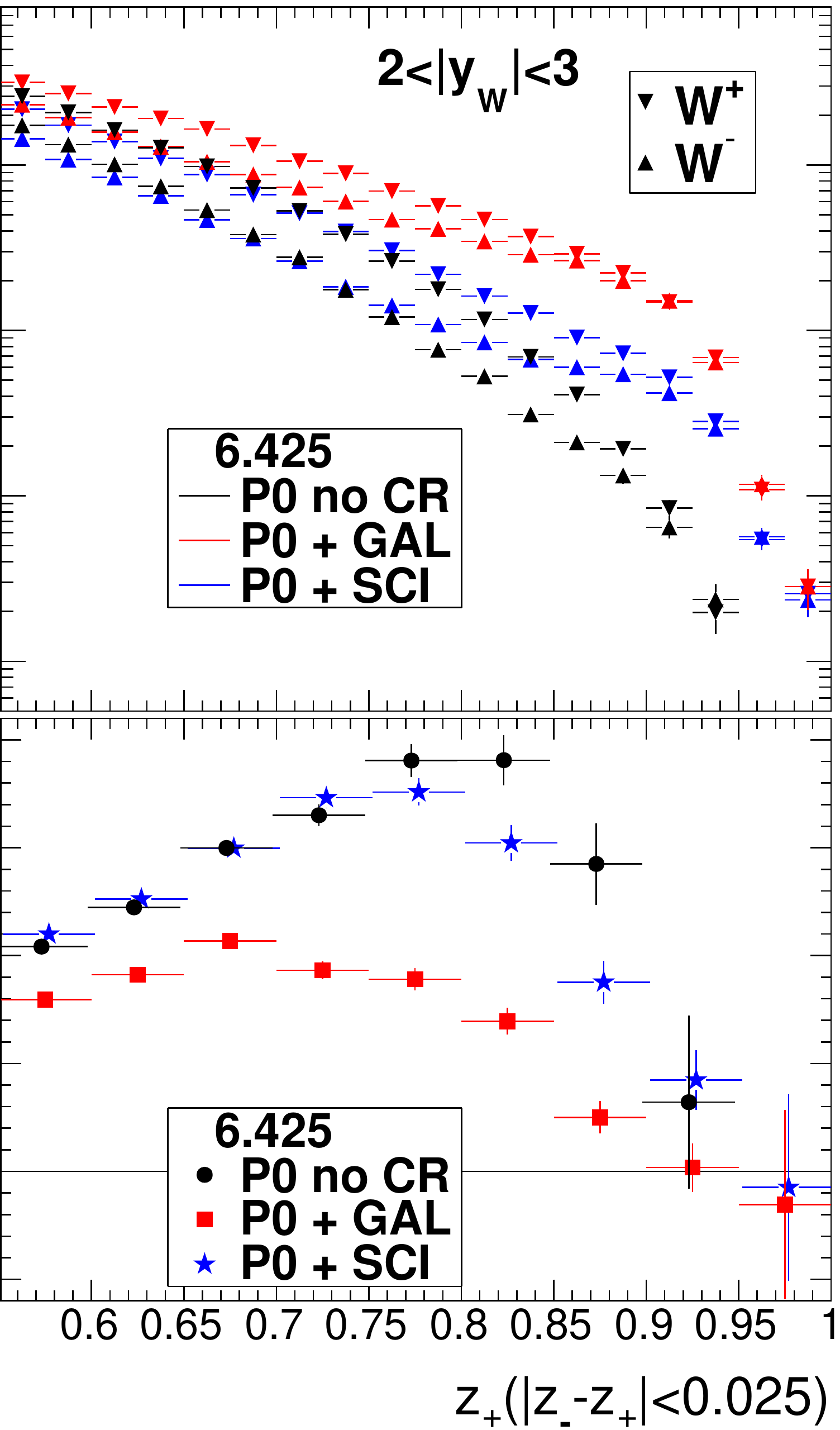}
\caption{ The differential cross sections in the
longitudinal momentum fraction of the leading proton moving in the
positive direction $z_{+}$ (with simultaneous requirement on the $z$
fraction of the second leading proton moving in the negative
direction $z_{-}$ as $|z_{-}-z_{+}|<0.025$), (top) and the
corresponding charge asymmetries (bottom) for different rapidity intervals as indicated.}
\label{fig:z-asym}
\end{figure*}

In addition to looking at the kinematics of the $W^\pm$'s produced
and the associated asymmetries, it is instructive to look at the
spectra of leading protons on both sides simultaneously. In order to
make the picture as clean as possible we show in
Fig.~\ref{fig:z-asym} the spectrum of protons in the positive
direction ($z_{+}$) when requiring a leading proton also on the negative
side ($z_{-}$) with similar momentum fraction $|z_{-}-z_{+}|<0.025$.
In addition we show the results not only for the GAL and SCI models
but also the results when neither of them is applied.

Similarly to the case of single leading protons, the characteristic
diffractive peak at $z\to 1$ can also be seen for the case with
double leading protons in Fig.~\ref{fig:z-asym} (top row).
However, it is visible at central $W$ rapidities only. For more
forward $W$ bosons the peak disappears, essentially due to momentum
conservation. Thus in order to obtain a selection of diffractive
events one has to apply also a cut on the rapidity of the W-bosons
in addition to the cuts on the leading protons. From the figure it is also
clear that the ``diffractive'' peak for central $W$'s is more pronounced
in the SCI case than in the GAL one. Similarly to the single leading proton case,
this is due to an increased production of leading protons for $z\gtrsim 0.6$
in the GAL model compared to the standard Perugia 0 tune,
whereas for the SCI model the additional double leading protons
are only seen for $z\gtrsim 0.85$.

Turning to the charge asymmetries we first note that in the limit $z\to 1$ the valence
quarks of the initial proton have to be part of the outgoing proton, so there
is no way to obtain any $W$ charge asymmetry in this case. Indeed,
in Fig.~\ref{fig:z-asym} (bottom row) we see the vanishing
asymmetry at large $z\to 1$ for both the GAL and SCI models.
At the same time,
since in the diquark fragmentation contribution both
valence and sea quarks may initiate the production of a
diffractive-like $W^{\pm}$, such a mechanism leads to a
noticeable $W$ charge asymmetry at moderate $z\lesssim 0.9$ (see
Fig.~\ref{fig:z-asym} -- bottom row).
From the figure it is also clear that the relative importance
of this contribution is larger for the SCI model than for the
GAL one giving larger asymmetries in the former case.
Finally, for larger $W$
rapidities the asymmetry is larger, which is due to an increasing
probability for a quark-initiated $W$ production.

Having studied the $W$ charge asymmetries in detail in both the GAL
and SCI models it is thus clear that the Monte Carlo simulation
affirms the statements made on general grounds, namely, the charge
asymmetry vanishes or, at least, becomes very small in the
asymptotic case $z \to 1$. Before coming to the conclusions we now
want to discuss the question of the origin of the asymmetry reported
recently in~\cite{GolecBiernat:2011dz}, where the earlier 6.215
version of {\sc Pythia} was used. Given the results obtained above
with {\sc Pythia} version 6.425 and the general arguments why there
should be no electric charge transfer in the $t$-channel in the
limit $z \to 1$, the observation of such an asymmetry may seem
contradictory. In order to be able to resolve this apparent
contradiction we have used the old 6.215 version of {\sc Pythia} in
the following. However, based on the observation made above that
there was no ``diffractive'' peak in the single leading proton
spectrum  when running the Monte Carlo with the same settings as
used by~\cite{GolecBiernat:2011dz} we have turned off the multiple
interactions.

\begin{figure}[!t]
\includegraphics[width=0.45\linewidth]{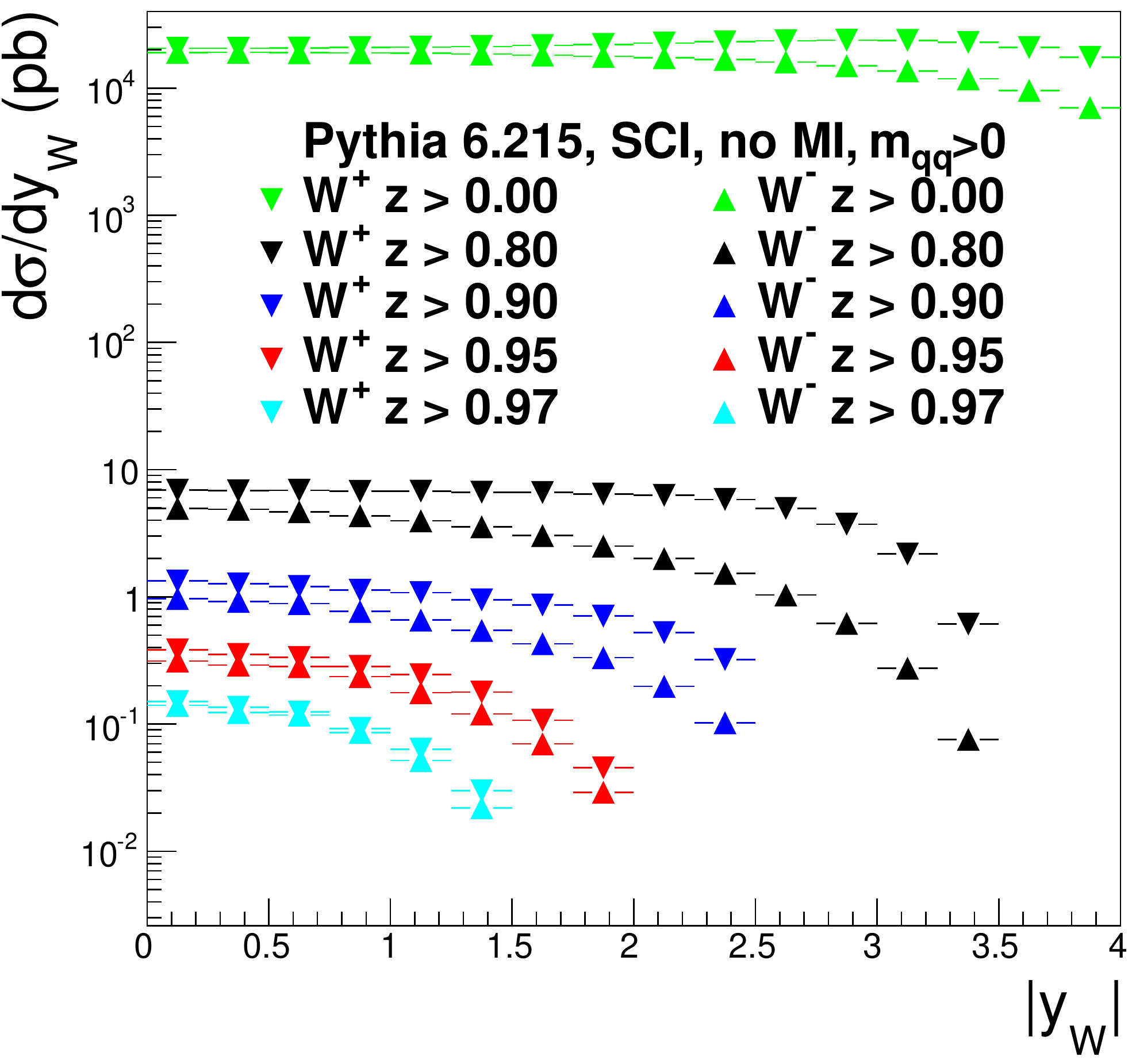}
\includegraphics[width=0.45\linewidth]{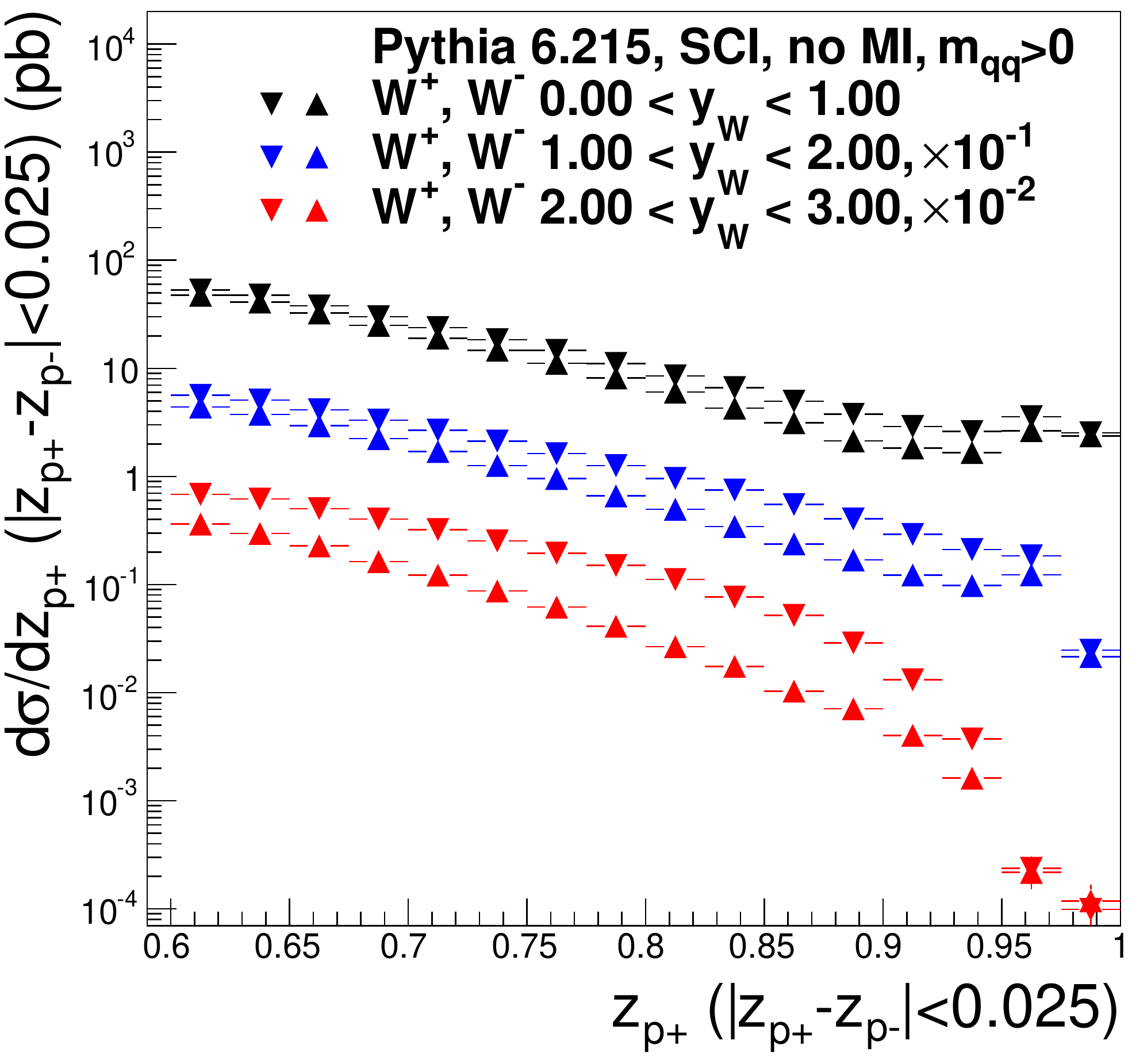}
\includegraphics[width=0.45\linewidth]{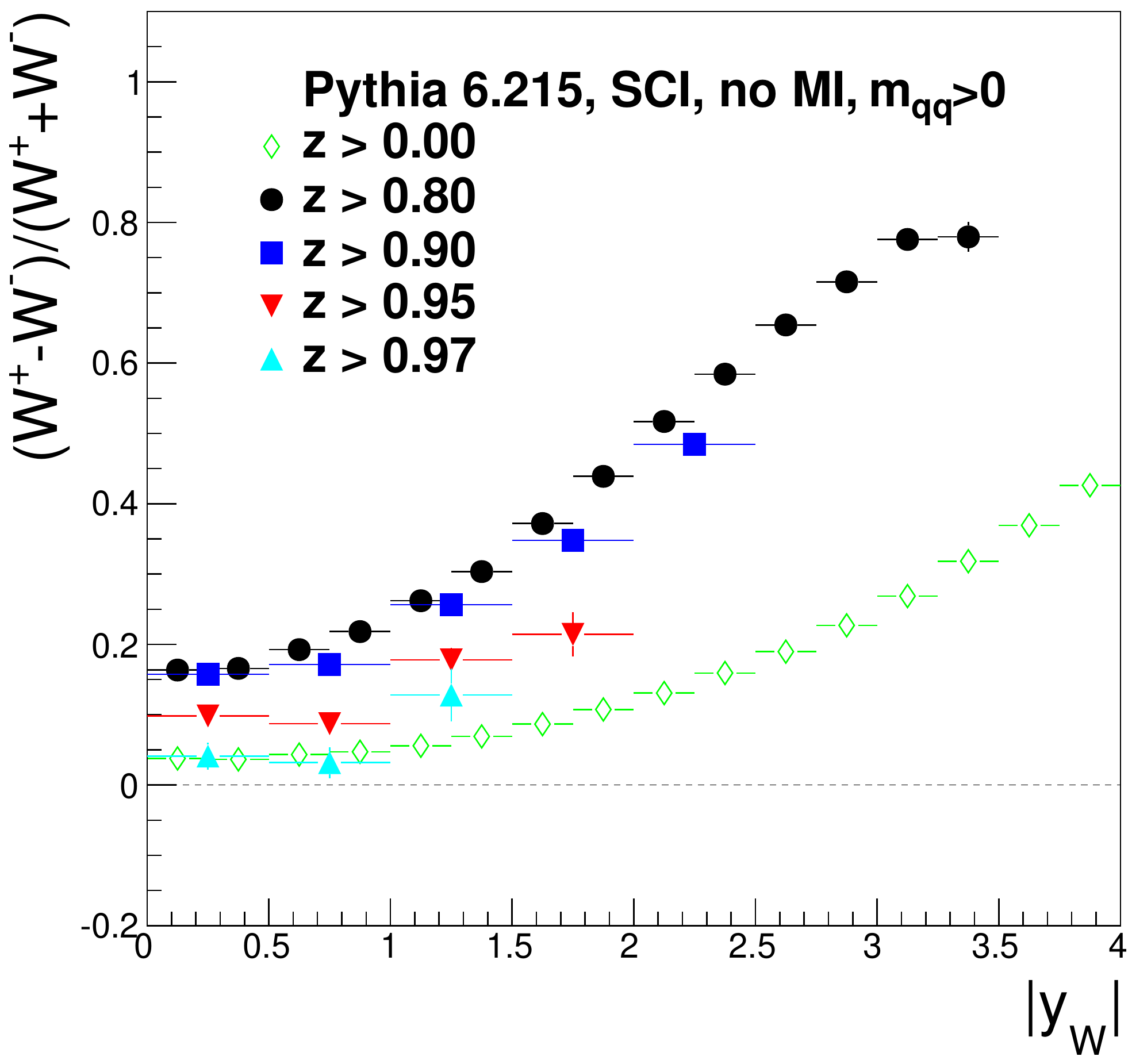}
\includegraphics[width=0.45\linewidth]{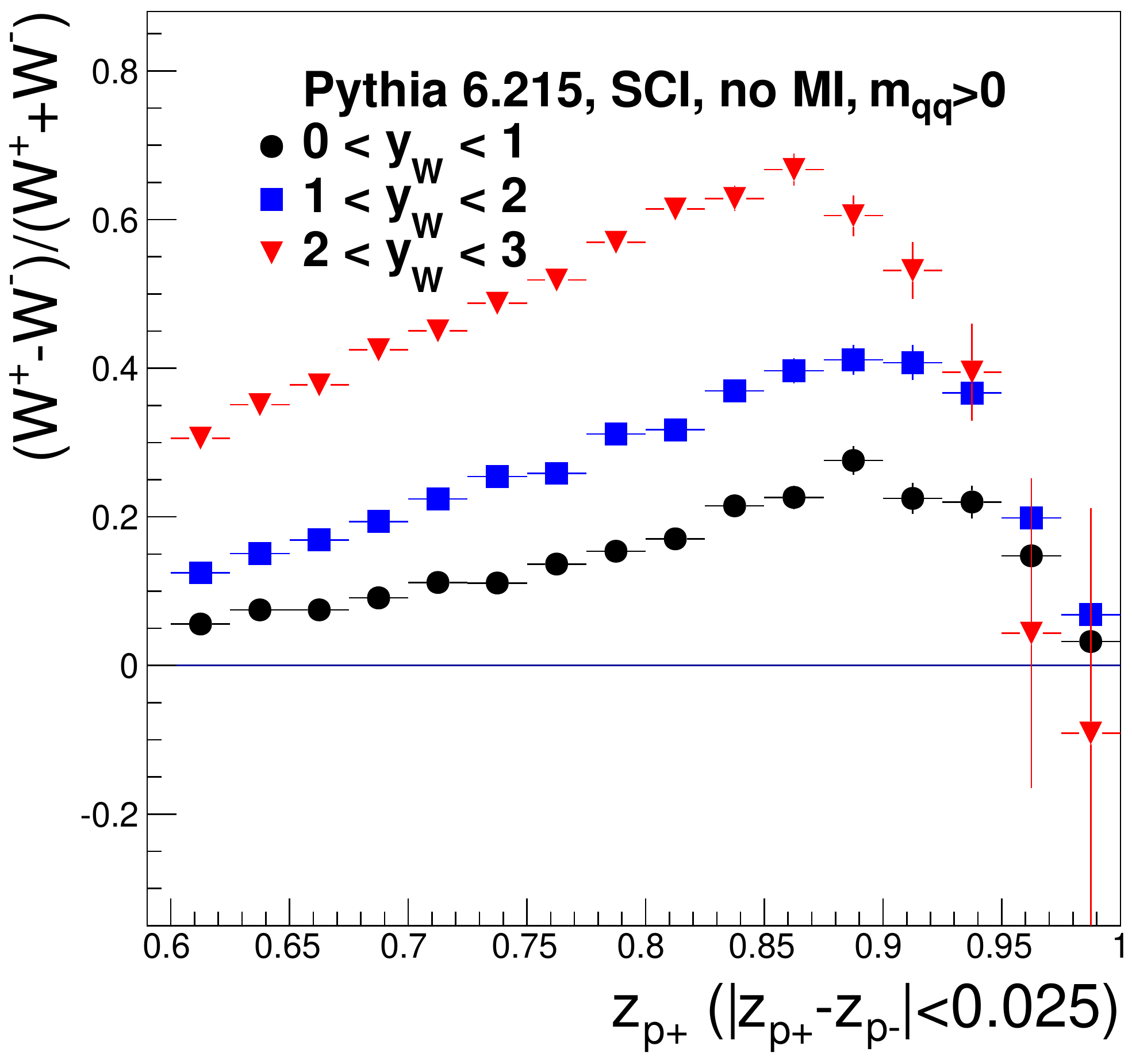}
\caption{
Diffractive $W$ production cross sections and $W$ charge asymmetry
when requiring both leading protons in earlier 6.215 version of {\sc
Pythia} with default settings but no multiple interactions.
Left: $d\sigma/dy_W$ for different
cuts on $\min (z^+,z^-)$ of leading protons.
Right: $d\sigma/dz_+$ of leading proton,
requiring both protons to have similar $z$, for
different bins in $W$ rapidity.
}
\label{fig:asym-py62}
\end{figure}

We start by investigating the cross-sections and corresponding asymmetries as
functions of the $W$ rapidity and the momentum fraction of the leading proton
on the positive side when requiring a leading proton on the negative side
with similar momentum as displayed in Fig.~\ref{fig:asym-py62}.
Comparing with the results obtained with {\sc Pythia 6.425}
there are two things that stand out. On the one hand the cross-sections
when requiring double leading protons are much smaller when using the old Pythia
version and the asymmetries are much larger. At the same time, in the limit $z \to 1$ it
is still the case that the asymmetries goes away. However, looking at the double
leading proton momentum fraction it is clear that even for central $W$'s there is
not really any diffractive-like peak in this case.

\begin{figure}[!t]
\includegraphics[width=0.45\linewidth]{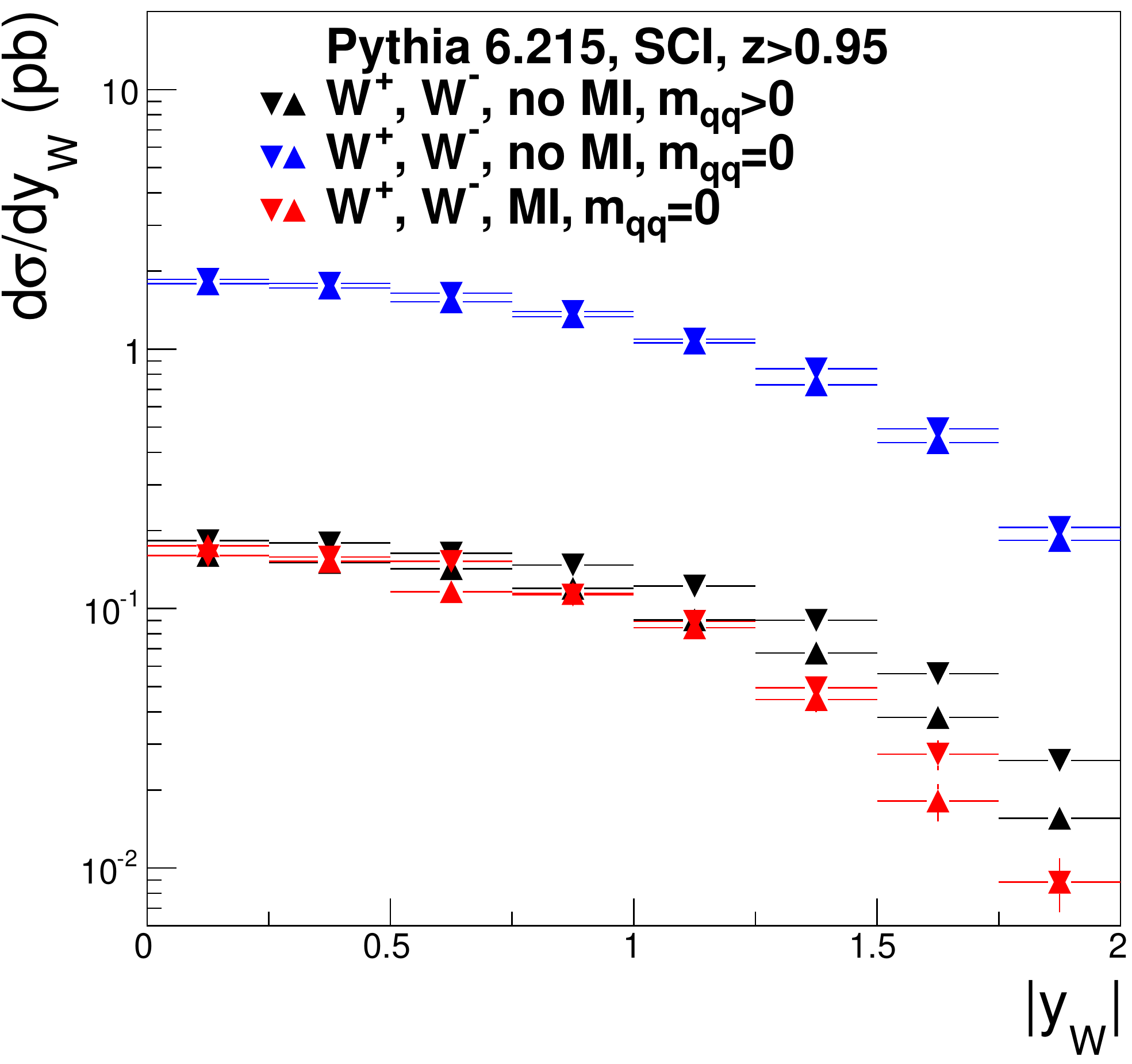}
\includegraphics[width=0.45\linewidth]{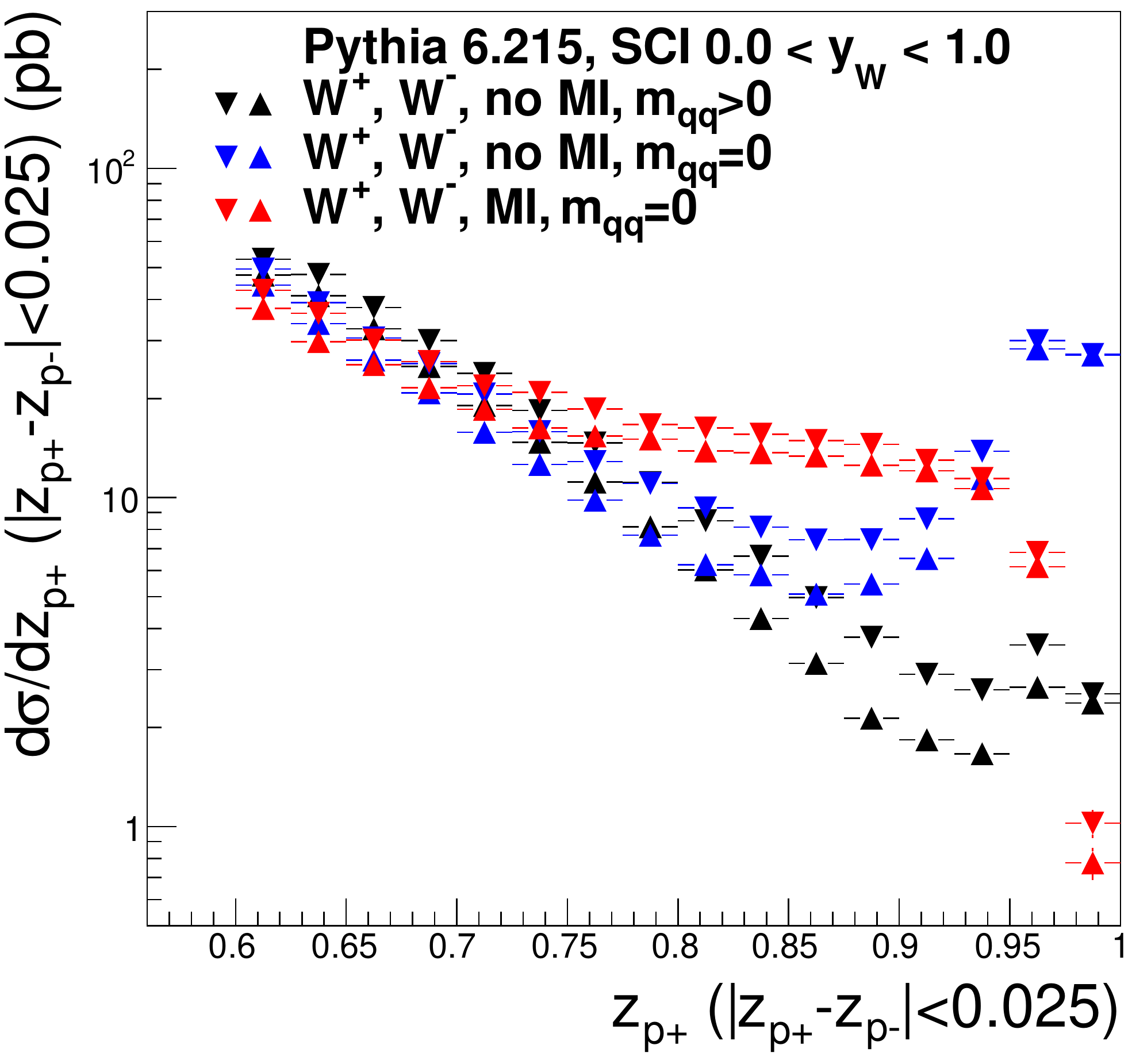}
\includegraphics[width=0.45\linewidth]{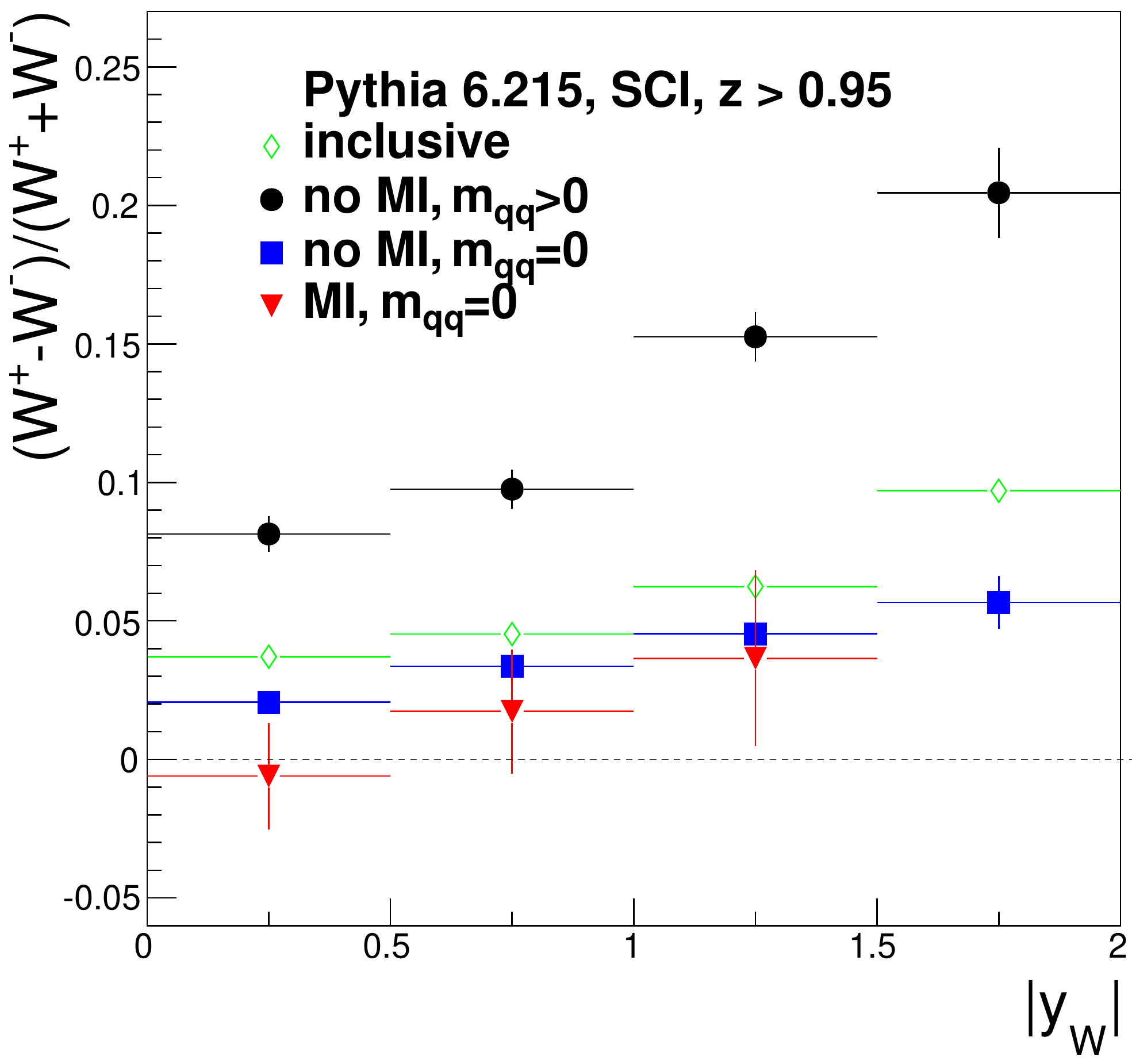}
\includegraphics[width=0.45\linewidth]{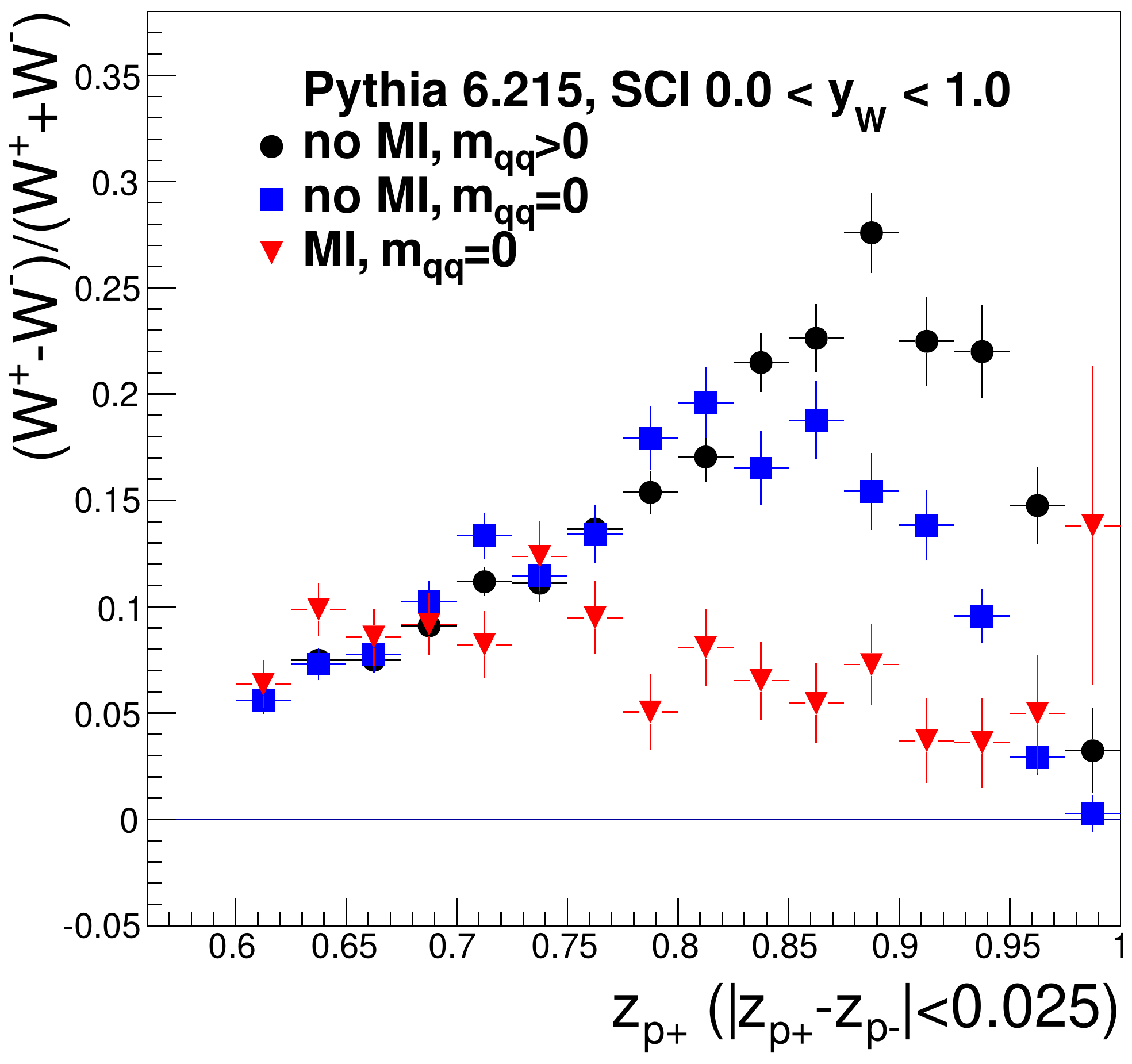}
\caption{
Comparison of the results for the cross-sections (top) and corresponding
charge asymmetries (bottom) as functions of the $W$ rapidity for $z>0.95$ (left)
and the momentum fraction $z$ for central rapidities $|y_W|<1$ (right)
comparing different choices of the diquark masses used in the remnant treatment,
as well as multiple interactions.
}
\label{fig:asym-py62-m0}
\end{figure}

The explanation of this apparently contradicting result has to do
with details of the Monte Carlo setup used
in~\cite{GolecBiernat:2011dz}. It has long been
known~\cite{Edin:1996mw} that the amount of leading protons depends
crucially on the constituent masses assigned in the Monte Carlo to
the valence quarks and diquarks in the proton remnant. The default
values in the 6.215 version are $m=0.33$ GeV for quarks and $m=0.58
\; (0.77)$ GeV for spin-0 (spin-1) diquarks. In addition the partons
in the proton remnant are given some transverse momentum. This means
that the invariant mass of the quark-diquark system will in most
cases be above the threshold for two-particle production such as
$p+\pi$. Then, most clusters will give two particles instead of only
one and hence very few high-$z$ protons ({\it cf.} the cluster
scaling factors in Fig.~\ref{fig:zcomp}).

In the later {\sc Pythia 6.425}, the kinematics of the remnant is calculated
using massless four-vectors for the valence quarks and diquarks.
This means that a much larger fraction of the clusters
will have invariant masses that are small enough to give just one proton (or other
baryon depending on the flavour and spin quantum numbers). To verify that this is
indeed the explanation, we show in
Fig.~\ref{fig:asym-py62-m0} the results obtained when setting the diquark masses to zero. (For clarity we only show the results for $z>0.95$.)
From the figure it is clear that this gives a substantial increase of the cross-section
and at the same time a decrease of the asymmetry.
Looking at the distribution in fractional momentum $z$
for centrally produced $W$'s
we see that with this setting there is a diffractive-like peak.
For reference we also show the results obtained when
including the multiple interactions.
This decreases both the magnitude of the cross-section and the asymmetries
but at the same time there is no diffractive peak, as demonstrated before.

\section{Summary and conclusions}

In this paper we have revisited the SCI and GAL colour reconnection
models for diffractive and exclusive $W^\pm X$ production, at LHC
($\sqrt{s}=14$ TeV) and Tevatron energies, when requiring both
single and double leading protons (or antiprotons for the Tevatron).
The requirement of a leading beam particle constitutes a much more
stringent test of the models than just requiring a rapidity gap and
also leads to sensitivity to other ingredients in the Monte Carlo,
in particular the constituent quark and diquark masses. Even so,
when applying the SCI and GAL models to the recent {\sc Pythia}
version 6.425 using the Perugia 0 tune, the resulting rates are in
overall agreement with data from the CDF experiment. Thus the models
can also be used to make predictions for the upcoming experiments at
the LHC implying, however, that there is an extra systematic
uncertainty related to the extrapolation from the Tevatron energy.

Looking at the spectra of both single as well as double leading
protons we see clear diffractive-like peaks for both the SCI and GAL
models. We have, however, shown that these peaks are sensitive to
other details of the Monte Carlo such as the amount of parton
showering, the implementation of multiple interactions, and the
constituent quark and diquark masses. Thus, in order to use these
models to make predictions for diffractive-like phenomena one  has
to take also these effects into account.

A focus of our paper has been on the issue of any possible $W$
charge asymmetry in diffractive $W^{\pm}X$ production with double
leading protons at the LHC. On general grounds, requiring two
leading protons there is no charge exchange in the t-channel, and
thus no charge asymmetry should exist. This is true for
diffractive-like events where the leading proton is produced from
the incoming beam with only a small momentum transfer. As we have
shown, in the Monte Carlo there are also other mechanisms, most
importantly diquark fragmentation, which may contribute more or less
to the amount of leading protons depending on how these are defined.
Based on our results we find that in both the SCI and GAL models the
diffractive-like protons starts to be significant when the outgoing
proton carries a fractional momentum $z$ of the beam energy which is
larger than $\sim0.9$ and only for $z\gtrsim 0.95$ do they dominate
the spectrum. In addition, for double leading protons, the
diffractive-like peak is only visible for centrally produced $W^\pm$
with rapidity $|y_W|\lesssim 1$.

Looking at $W$'s produced centrally and with double leading protons
each having $z>0.9$ we find that the charge asymmetry, present when
looking inclusively, goes away at the percent level in agreement
with the general expectations. Even so there are details that differ
between the two colour reconnection models. Fig.~\ref{fig:zcomp}
shows that both have the same shape of the diffractive peak, the
main difference is that the underlying background level of the
proton $z$-spectrum is higher for the GAL model. This difference is
also seen in Fig.~\ref{fig:z-asym}. In addition, the charge
asymmetry is smaller in the GAL model -- going to zero around
$z\sim0.8$, whereas in the SCI model the asymmetry is close to or
larger than the inclusive one for $z\lesssim 0.8$. Thus, in this
non-diffractive region, the charge asymmetry and double leading
proton spectrum can potentially be used to discriminate between the
SCI and GAL models.

Finally we have clarified that the charge asymmetry observed
in~\cite{GolecBiernat:2011dz} originates in the use of the older
{\sc Pythia 6.215} multiple interactions model, default constituent
quark and diquark masses and a leading proton definition requiring
the relaxed cut $z>0.85$. As a consequence the fraction of
diffractive like protons is very small and instead the results are
completely dominated by the diquark fragmentation contribution,
making the result incompatible with a pomeron-based model which does
inherently only describe the diffractive part.

A major strength of the colour exchange models, such as SCI and GAL,
is that they describe both diffractive and inclusive events with a
smooth transition in-between. The GAL model is based on a
string-field minimization property that may reveal important aspects
of the soft QCD colour field. The SCI model has recently been
developed into a proper QCD-based model \cite{Pasechnik:2010zs} for
diffractive deep inelastic scattering that does describe the salient
features of data from HERA. Since this model is derived from $k_T$
factorisation at the amplitude level it is non-trivial to cast into
a probabilistic Monte Carlo framework, but such an extension is
under development \cite{DDISSCI:2013} in order to facilitate more
detailed comparisons with data. Models of the kind studied in detail
in this paper will be tested by the expected LHC data on various
diffractive processes, which should increase our understanding of
soft QCD dynamics.

\vspace{5mm} \acknowledgments We thank Otto Nachtmann, Christophe
Royon, Torbj\"orn Sj\"ostrand and Peter Skands for valuable
discussions and Stefan Prestel for technical help. We also thank
David Eriksson and Oscar St{\aa}l for sharing their implementation
of the SCI and GAL models in {\sc Pythia} 6.4. This work is
supported in part by the Swedish Research Council grants
621-2011-5333 and 621-2011-5107.


\end{document}